\documentclass[aps,prd,preprintnumbers,superscriptaddress,nofootinbib,notitlepage]{revtex4-2}
\usepackage[pdftex]{graphicx}
\usepackage{bm,latexsym,amsmath,amssymb,amsfonts,mathrsfs}
\usepackage{mathtools}
\usepackage{color}
\usepackage{float}
\usepackage{subcaption} 
\allowdisplaybreaks[1]
\usepackage[pdftex,colorlinks=true,linkcolor=blue,citecolor=cyan]{hyperref}
\begin{document}
	\title{Chaotic motion of spinning particles in nonlocal gravity}

	\author{Jianhui Lin}
	\email[Email: ]{202510188472@mail.scut.edu.cn}
	\affiliation{Department of Physics, South China University of Technology, Guangzhou 510641, China.}
	%
	
	%\author{Yunlong Liu}
	%\email[Email: ]{phliuyunlong@kust.edu.cn}
	%\affiliation{School of Science, Kunming University of Science and Technology, Kunming 650093, China.}
	%
	
	\author{Xiangdong Zhang}
	\email[Corresponding author: ]{scxdzhang@scut.edu.cn}
	\affiliation{Department of Physics, South China University of Technology, Guangzhou 510641, China.}

	\begin{abstract}
		We study spin-induced chaos around a static, spherically symmetric black hole in revised Deser--Woodard nonlocal gravity (NLG), using the conservative test-body approximation relevant to extreme-mass-ratio inspirals. The deformation parameter $\zeta$ describes departures from Schwarzschild spacetime. Poincar\'e sections, rotation curves, and fast Lyapunov indicators (FLIs) reveal changes in resonant structures and a decreasing fraction of non-plunging trajectories classified as chaotic within the sampled family, from $37.5\%$ at $\zeta=0$ to zero at $\zeta=0.1$. The FLI classifications of selected trajectories are reflected in their approximate mass-quadrupole waveforms: neighboring chaotic and regular trajectories show a marked local spectral-flatness contrast, which weakens in backgrounds where both are classified as regular. To assess whether a transition from chaotic to regular motion caused by the background deformation helps distinguish NLG from GR, we compare NLG waveforms at $\zeta=10^{-6}$ with GR waveforms generated at different initial radii. For a trajectory whose classification changes between the backgrounds, this adjustment substantially reduces the mismatch. The selected NLG waveforms from chaotic orbits retain larger minimum mismatches than those from regular orbits, although sensitivity to initial conditions may also hinder signal recovery. These results connect background-dependent orbital dynamics with radiation spectra, but do not establish that chaos offers an advantage for identifying the background geometry.
	\end{abstract}

	%-----------------------------------------------------------------%
	\maketitle
	%-----------------------------------------------------------------%
	%%%%%%%%%%%%%%%%%%%%%%%%%%%%%%%%%%%%%%%%%%%%%%%%%%%%%%%%%%%%%%%%%%%%%%%%%%%%%%%%%%%%%%%%%%%%%%%%%%%%%%%%%%%%%%%%%%%%%%%
	\section{Introduction}
	\label{sec:intro}
	
	The spin of a compact body affects its motion around a black hole through its coupling to spacetime curvature. This interaction is relevant to extreme-mass-ratio inspirals (EMRIs), in which a stellar-mass compact object gradually spirals into a supermassive black hole \cite{AmaroSeoaneEtAl2007,AmaroSeoane2018}. The gravitational waves emitted by these systems carry information about orbital dynamics and spacetime geometry in the strong-field region, making EMRIs important sources for space-based detectors such as LISA \cite{BabakEtAl2017,GairEtAl2013}. In the conservative test-body approximation, the secondary can be modeled as a spinning body described by its mass monopole and spin dipole. Neglecting higher multipole moments, its motion is governed by the Mathisson--Papapetrou--Dixon (MPD) equations \cite{Mathisson1937,Papapetrou1951,Dixon1970,Dixon1974}. In Schwarzschild and Kerr spacetimes, spinless geodesic motion admits enough conserved quantities to be integrable. Spin--curvature coupling modifies this dynamics, allowing spinning particles to exhibit resonances and chaos even when geodesic motion in the same background is integrable \cite{SuzukiMaeda1997,Han2008,ZelenkaEtAl2020}.
	
	Chaotic motion of spinning particles in Schwarzschild spacetime was demonstrated using Poincar\'e sections and Lyapunov exponents \cite{SuzukiMaeda1997}. The accompanying effective-potential analysis also linked changes in the orbital structure to the particle spin. Later studies carried out systematic searches for chaotic trajectories in Kerr spacetime \cite{Hartl2003,Hartl2003b}. Although many chaotic solutions were found, pronounced chaos generally required spins much larger than those expected in EMRIs. 
	
	Chaotic orbits with astrophysically attainable spins were reported in Schwarzschild spacetime \cite{VerhaarenHirschmann2010}. The astrophysical relevance of this spin-induced chaos was questioned. A subsequent comment pointed out theoretical and technical issues concerning the normalization of mass and spin and the interpretation of Poincar\'e sections and Lyapunov exponents \cite{LukesGerakopoulos2016Comment}. A separate study of a linear-in-spin Hamiltonian approximation in Kerr spacetime found that the effects of nonintegrability on the motion appeared negligible in the astrophysically relevant small-spin regime examined \cite{LukesGerakopoulosEtAl2016}.
	
	Later studies identified small chaotic regions at astrophysically relevant spin values. An analysis of resonances in Schwarzschild spacetime found that the resonances examined were generated by terms quadratic in the particle spin \cite{ZelenkaEtAl2020}. Their widths become smaller as the spin decreases, making these structures difficult to resolve at small spins. Subsequent work examined approximate radiation spectra for selected neighboring regular and chaotic trajectories \cite{YuanEtAl2026}. Their time-domain signals can appear similar, while the selected chaotic waveforms contain more spectral power between the dominant peaks. Local spectral flatness was used to quantify this contrast.
	
	These studies motivate an examination of how spin-induced resonances and chaos respond to changes in the background geometry. A deformation of Schwarzschild spacetime changes the curvature entering the spin--curvature coupling and may alter the nearby orbital structures. Previous work in a Horndeski black-hole background has examined the effects of scalar hair on spinning-particle dynamics and approximate waveforms \cite{YuEtAl2026}. Here we study this question in revised Deser--Woodard nonlocal gravity (NLG). The theory modifies the gravitational action through nonlocal terms to account for cosmic acceleration without introducing a cosmological constant \cite{DeserWoodard2019}. These terms contain inverse differential operators acting on curvature quantities, so the field equations depend on more than the local curvature alone. Static, spherically symmetric black-hole solutions have also been constructed, providing backgrounds for studying the strong-field effects of the theory \cite{DW2025}. Many aspects of NLG have been investigated under this static spherically symmetric black-hole solutions \cite{LiZhang2026Lensing,BravoGaeteEtAl2026,LiZhang2026Extended,DA2026,DAgostinoDeFalcoQNM2025,DAgostinoDeFalcoLensing2026,LiuDu2026}.
	
	The NLG solution considered here contains a deformation parameter $\zeta$. It reduces to Schwarzschild spacetime as $\zeta\to0$ and remains static and spherically symmetric for nonzero $\zeta$. Spinless geodesic motion therefore remains integrable throughout this family. This condition allows us to examine how modifications of the background geometry affect spin‑dependent resonances and chaos, without introducing additional sources of non‑integrability that may arise from a non‑spherical background. We study whether the deformation merely shifts the resonant structures or also changes the occurrence of finite-time chaos within a chosen family of initial conditions, and whether these changes are reflected in approximate radiation spectra.
	
	We further ask whether a transition from chaotic to regular motion caused by the background deformation can help distinguish NLG from GR. Since spinning-particle motion can be chaotic in both backgrounds, the presence of chaos alone does not distinguish them. A change in dynamical classification may be accompanied by waveform changes, but adjusting the orbital initial conditions may substantially reduce the waveform differences between the backgrounds. We therefore compare selected NLG waveforms with a common set of GR waveforms generated at different initial radii. This comparison examines whether the differences accompanying the classification change remain appreciable after the GR initial radius is adjusted, and whether NLG waveforms from chaotic orbits retain larger minimum mismatches than those from regular orbits.
	
    This paper is organized as follows. In Sec.~\ref{sec:spinning particle}, we first introduce the black-hole background in revised Deser--Woodard NLG. We then describe the MPD equations and specify the initial conditions using the effective potential. In Sec.~\ref{sec:diagnostics}, we examine the orbital dynamics through Poincar\'e sections, rotation curves, and fast Lyapunov indicators, comparing the resonant structures and the distribution of chaotic trajectories for different values of $\zeta$. In Sec.~\ref{sec:waveforms}, we construct approximate gravitational waveforms and compare their time-domain signals, Fourier spectra, and local spectral flatness. We then examine noise-weighted waveform mismatches and their dependence on the GR initial radius. Finally, Sec.~\ref{sec:conclusions} summarizes the results and discusses their limitations and possible extensions.
	
	%%%%%%%%%%%%%%%%%%%%%%%%%%%%%%%%%%%%%%%%%%%%%%%%%%%%%%%%%%%%%%%%%%%%%%%%%%%%%%%%%%%%%%%%%%%%%%%%%%%%%%%%%%%%%%%%%%%%%%%
	\section{Spinning particle in nonlocal gravity}
	\label{sec:spinning particle}
	%%%%%%%%%%%%%%%%%%%%%%%%%%%%%%%%%%%%%%%%%%%%%%%%%%%%%%%%%%%%%%%%%%%%%%%%%%%%%%%%%%%%%%%%%%%%%%%%%%%%%%%%%%%%%%%%%%%%%%%
	
	%%%%%%%%%%%%%%%%%%%%%%%%%%%%%%%%%%%%%%%%%%%%%%%%%%%%%%%%%%%%%%%%%%%%%%%%%%%%%%%%%%%%%%%%%%%%%%%%%%%%%%%%%%%%%%%%%%%%%%%
	\subsection{Black hole solutions in nonlocal gravity}
	\label{sec:DW_black_hole}
	%%%%%%%%%%%%%%%%%%%%%%%%%%%%%%%%%%%%%%%%%%%%%%%%%%%%%%%%%%%%%%%%%%%%%%%%%%%%%%%%%%%%%%%%%%%%%%%%%%%%%%%%%%%%%%%%%%%%%%%
	
	Nonlocal gravity modifies general relativity by introducing nonlocal operators into the gravitational action. These modifications have been studied in connection with quantum effects and cosmological evolution, particularly cosmic acceleration \cite{CapozzielloBajardi2022}. Two main approaches are infinite-derivative theories and integral-kernel theories. Infinite-derivative theories involve nonpolynomial analytic functions of the covariant d'Alembertian $\Box$, whose expansion contains infinitely many derivative terms. These theories have been developed to address ultraviolet divergences and spacetime singularities, with suitable models admitting nonsingular gravitational solutions. Integral-kernel theories instead introduce operators such as $\Box^{-1}$ acting on curvature quantities. This inverse operator can be written in integral form, relating fields at different spacetime points and making the nonlocal dependence explicit. Such models have been explored as descriptions of infrared gravitational corrections and as alternatives to dark energy in cosmology. The Deser--Woodard model and its revised formulation belong to this class of nonlocal theories \cite{DeserWoodard2019}.
	
	We consider the static, spherically symmetric black hole solution constructed by D'Agostino and De Falco in revised Deser--Woodard NLG \cite{DW2025}. Its gravitational action reads
	\begin{equation}
		S=\frac{1}{16\pi G}
		\int d^{4}x\,\sqrt{-g}\,R[1+F(Y)],
		\label{DW_action}
	\end{equation}
	where $F(Y)$ is the distortion function and
	\begin{equation}
		X=\Box^{-1}R,
		\qquad
		Y=\Box^{-1}\left(g^{ab}\nabla_aX\nabla_bX\right),
		\label{DW_scalars}
	\end{equation}
	with $\Box=g^{ab}\nabla_a\nabla_b$. The inverse d'Alembertian is defined through a retarded Green function, so that the fields at a given event depend on sources within its causal past. This nonlocal modification is motivated by a conjecture about the cosmologically significant contributions to the quantum gravitational effective action. The distortion function $F(Y)$ is not derived from an underlying quantum theory.
	
	Introducing auxiliary fields $U$ and $V$, the nonlocal field equations can be written in local differential form,
	\begin{subequations}
		\label{DW_auxiliary}
		\begin{align}
			\Box X &= R,\\
			\Box Y &= g^{ab}\nabla_aX\nabla_bX,\\
			\Box U &= -2\nabla_a(V\nabla^aX),\\
			\Box V &= R\frac{dF}{dY}.
		\end{align}
	\end{subequations}
	The local equations alone do not fix the homogeneous solutions of the auxiliary fields. To recover the original nonlocal model, their solutions must follow the prescription defining $\Box^{-1}$. In the cosmological formulation of Ref.~\cite{DeserWoodard2019}, this is a retarded prescription, with the auxiliary fields and their first normal derivatives vanishing on an initial spacelike hypersurface. Defining
	\begin{equation}
		K_{ab}
		=\nabla_aX\nabla_bU+\nabla_aY\nabla_bV
		+V\nabla_aX\nabla_bX,
	\end{equation}
	the modified vacuum field equations are
	\begin{equation}
		\left(G_{ab}+g_{ab}\Box-\nabla_a\nabla_b\right)
		\left[1+U+F(Y)\right]
		+K_{(ab)}
		-\frac{1}{2}g_{ab}g^{cd}K_{cd}=0.
		\label{DW_field_equations}
	\end{equation}
	The terms involving $U$, $F(Y)$, and $K_{ab}$ encode the nonlocal corrections to the Einstein equations.
	
	To construct static, spherically symmetric black hole solutions, we adopt the metric ansatz
	\begin{equation}
		ds^2=-A(r)dt^2+\frac{dr^2}{B(r)}
		+r^2(d\theta^2+\sin^2\theta\,d\phi^2),
		\label{DW_metric}
	\end{equation}
	where $A(r)$ and $B(r)$ are radial functions to be determined from the field equations. Following Ref.~\cite{DW2025}, we adopt $G=c=M=1$, with coordinates normalized by the black hole mass scale $M$. The construction requires asymptotic flatness, positive and smooth metric functions outside the horizon. By expressing the field equations in a static orthonormal frame and solving them perturbatively around the Schwarzschild geometry, one obtains, to first order in the deformation parameter $\zeta$,
	\begin{subequations}
		\label{DW_solution}
		\begin{align}
			A(r)&=1-\frac{2}{r}-\frac{\zeta}{r^n},
			\label{DW_A}\\
			B(r)&=1-\frac{2}{r}
			+\frac{\zeta}{3^nr^{n+1}(r-3)^2}
			\Bigl\{
			3^nr[n(r-3)(r-2)+4r-9]
			\nonumber\\
			&\hspace{35mm}
			-3(r-2)(2r-3)r^n
			\Bigr\}.
			\label{DW_B}
		\end{align}
	\end{subequations}
	The apparent singularity at $r=3$ is removable. The dimensionless parameters $\zeta$ and $n$ characterize the deformation amplitude and the inverse-power index, respectively. The black hole solution considered here satisfies
	\begin{equation}
		0<\zeta\ll1,
		\qquad n>1.
	\end{equation}
	The perturbative construction requires a small deformation parameter but does not specify a numerical upper bound on $\zeta$. Previous studies of this black hole geometry have considered deformation parameters up to $0.1$, including analyses of the background solution, quasinormal modes, and gravitational lensing \cite{DW2025,DAgostinoDeFalcoQNM2025,DAgostinoDeFalcoLensing2026}. Following these choices, we restrict our numerical study to $0\leq\zeta\leq0.1$ and fix $n=2$ throughout. The value $\zeta=0.1$ is the upper limit of our parameter scan rather than a theoretical or observational bound. The Schwarzschild geometry is recovered for $\zeta=0$. To first order, the event horizon is located at
	\begin{equation}
		r_{\mathrm H}
		=2+\frac{\zeta}{2^{n-1}}
		+\mathcal{O}(\zeta^2).
	\end{equation}
	
	Finally, the distortion function is reconstructed from the auxiliary-field equations~\eqref{DW_auxiliary}. Once the metric is known, to first order in $\zeta$, this procedure gives
	\begin{equation}
		F(Y)
		=\zeta\frac{3^{1-n}-[r(Y)]^{1-n}}{3-r(Y)},
		\qquad
		r(Y)=\frac{2}{1-e^{-Y/\zeta}}.
		\label{DW_distortion}
	\end{equation}
	This reconstructed function specifies the distortion term in Eq.~\eqref{DW_action} associated with the black hole geometry to the perturbative order considered.
	
	%%%%%%%%%%%%%%%%%%%%%%%%%%%%%%%%%%%%%%%%%%%%%%%%%%%%%%%%%%%%%%%%%%%%%%%%%%%%%%%%%%%%%%%%%%%%%%%%%%%%%%%%%%%%%%%%%%%%%%%
	\subsection{Equations of motion for spinning particles}
	\label{sec:Eq of spinning particles}
	%%%%%%%%%%%%%%%%%%%%%%%%%%%%%%%%%%%%%%%%%%%%%%%%%%%%%%%%%%%%%%%%%%%%%%%%%%%%%%%%%%%%%%%%%%%%%%%%%%%%%%%%%%%%%%%%%%%%%%%
	
	We consider the conservative test-body limit of an extreme mass-ratio inspiral (EMRI) system, neglecting gravitational self-force effects. The secondary is modeled as a spinning extended test body whose mass is much smaller than that of the central compact object. Its motion takes place in a fixed background spacetime generated by the central object. Within Dixon's covariant multipole formalism, truncating the multipole expansion at the pole--dipole order, with quadrupole and higher-order moments neglected, yields the Mathisson--Papapetrou--Dixon (MPD) equations,
	\begin{align}
		\frac{DP^{a}}{D\tau}
		&= -\frac{1}{2}R^{a}{}_{bcd}u^{b}S^{cd},
		\label{MPD1}\\
		\frac{DS^{ab}}{D\tau}
		&= P^{a}u^{b}-P^{b}u^{a}.
		\label{MPD2}
	\end{align}
	Here, $\tau$ denotes the proper time along the representative worldline, $u^{a}=dx^{a}/d\tau$ is its four-velocity, and $D/D\tau=u^{a}\nabla_{a}$ denotes the covariant derivative along this worldline. We adopt the metric signature $(-,+,+,+)$, so that $u^{a}u_{a}=-1$. The quantities $P^{a}$ and $S^{ab}=-S^{ba}$ are the four-momentum and spin angular momentum tensor of the particle, respectively, and $R^{a}{}_{bcd}$ is the Riemann curvature tensor of the background spacetime. The right-hand side of Eq.~\eqref{MPD1} describes the coupling between the particle's spin and the background curvature.
	
	The MPD equations alone do not form a closed system determining the motion of the spinning particle. This underdetermination reflects the freedom to choose the representative worldline within the extended body, relative to which its multipole moments are defined. A spin supplementary condition (SSC) is therefore required to select a representative centroid and its associated worldline. In this work, we adopt the Tulczyjew--Dixon (TD) condition \cite{Tulczyjew1959,Dixon1970},
	\begin{equation}
		S^{ab}P_{b}=0.
		\label{TD}
	\end{equation}
	This condition selects the particle's center of mass in its momentum rest frame, where the mass dipole moment vanishes. Together with the proper-time normalization, it determines the relation between the four-velocity and the four-momentum and closes the MPD system.
	
	Under the TD condition, the dynamical mass $\mu$ and the spin magnitude $s$, defined by
	\begin{equation}
		\mu^{2}=-P^{a}P_{a},
		\qquad
		s^{2}=\frac{1}{2}S^{ab}S_{ab},
		\label{mass_spin}
	\end{equation}
	are conserved along the motion within the pole--dipole approximation. In general, the four-velocity $u^{a}$ and the four-momentum $P^{a}$ are not parallel owing to spin--curvature coupling. More explicitly, combining the MPD equations with the TD condition yields \cite{CostaNatario2015}
	\begin{equation}
		u^{a}=\frac{dx^{a}}{d\tau}
		=\frac{m}{\mu^{2}}
		\left(
		P^{a}
		+\frac{2S^{ab}R_{bcde}P^{c}S^{de}}
		{4\mu^{2}+R_{bcde}S^{bc}S^{de}}
		\right),
		\label{velocity_momentum}
	\end{equation}
	where $m=-P^{a}u_{a}$ is the mass with respect to the four-velocity and is generally not conserved. The proper-time normalization gives
	\begin{equation}
		m=\frac{\mathcal{A}\mu^{2}}
		{\sqrt{\mathcal{A}^{2}\mu^{2}-\mathcal{B}s^{2}}},
		\label{mass_normalization}
	\end{equation}
	where
	\begin{subequations}
		\label{normalization_definitions}
		\begin{align}
			\mathcal{A}
			&=4\mu^{2}+R_{abcd}S^{ab}S^{cd},
			\label{normalization_A}\\
			\mathcal{B}
			&=4\mathcal{D}^{ab}R_{acde}P^{c}S^{de}
			R_{bfgh}P^{f}S^{gh},
			\label{normalization_B}\\
			\mathcal{D}^{a}{}_{b}
			&=\frac{1}{s^{2}}S^{ac}S_{bc}.
			\label{normalization_h}
		\end{align}
	\end{subequations}
	Substituting Eq.~\eqref{mass_normalization} into Eq.~\eqref{velocity_momentum} expresses the four-velocity entirely in terms of the four-momentum, spin tensor, and background curvature. The evolution equations for $x^{a}$, $P^{a}$, and $S^{ab}$ therefore form a closed system.
	
	%%%%%%%%%%%%%%%%%%%%%%%%%%%%%%%%%%%%%%%%%%%%%%%%%%%%%%%%%%%%%%%%%%%%%%%%%%%%%%%%%%%%%%%%%%%%%%%%%%%%%%%%%%%%%%%%%%%%%%%
	\subsection{Integrability of nonspinning and spinning particle motion}
	\label{sec:integrability}
	%%%%%%%%%%%%%%%%%%%%%%%%%%%%%%%%%%%%%%%%%%%%%%%%%%%%%%%%%%%%%%%%%%%%%%%%%%%%%%%%%%%%%%%%%%%%%%%%%%%%%%%%%%%%%%%%%%%%%%%
	According to the Liouville--Arnold theorem, a Hamiltonian system with $N$ degrees of freedom is Liouville integrable if it admits $N$ functionally independent integrals of motion, including the Hamiltonian, that are pairwise in involution. We use this criterion to compare nonspinning and spinning particle motion in the static, spherically symmetric background considered here.
	
	The Hamiltonian description of nonspinning-particle motion and its integrability are discussed in Refs.~\cite{FrolovEtAl2017,RiosecoSarbach2017,ZhangJiang2025}. For the static, spherically symmetric metric~\eqref{DW_metric}, timelike geodesic motion is governed by
	\begin{equation}
		H_{\mathrm{geo}}
		=\frac{1}{2}g^{ab}P_aP_b=-\frac{\mu^2}{2},
	\end{equation}
	which provides one integral of motion. The covariant phase space is eight-dimensional, with four coordinate--momentum pairs $(x^a,P_a)$. Time-translation and rotational symmetries give three additional integrals,
	\begin{equation}
		E=-P_t,
		\qquad
		L_z=P_\phi,
		\qquad
		L^2=P_\theta^2+\frac{P_\phi^2}{\sin^2\theta},
	\end{equation}
	where $L_z$ is the angular momentum component along the polar axis and $L^2$ is the squared total angular momentum. Together with $H_{\mathrm{geo}}$, these quantities are functionally independent at generic phase-space points and mutually Poisson commuting. They form the complete set of four integrals required for Liouville integrability, independently of the particular radial functions $A(r)$ and $B(r)$.
	
	A Hamiltonian formulation of the full MPD--TD equations is given in Ref.~\cite{WitzanyEtAl2019}. For a spinning particle, the TD supplementary condition and the fixed spin magnitude leave only one additional degree of freedom \cite{RamondIsoyama2026}. In contrast to the geodesic case, the phase-space dimension therefore increases from eight to ten, and the number of degrees of freedom increases from four to five. The static and spherical symmetries of the spacetime give rise to conserved quantities. For a Killing vector field $\xi^a$, the corresponding integral is
	\begin{equation}
		C(\xi^a)=P_a\xi^a+\frac{1}{2}S^{ab}\nabla_a\xi_b.
	\end{equation}
	The timelike Killing vector field $\xi_{(t)}^a=(\partial_t)^a$ generates the conserved energy $E=-C(\xi_{(t)}^a)$, while the three spacelike Killing vector fields
	\begin{align}
		\xi_{(x)}^a
		&=-\sin\phi\,(\partial_\theta)^a
		-\cot\theta\cos\phi\,(\partial_\phi)^a,\\
		\xi_{(y)}^a
		&=\cos\phi\,(\partial_\theta)^a
		-\cot\theta\sin\phi\,(\partial_\phi)^a,\\
		\xi_{(z)}^a
		&=(\partial_\phi)^a,
	\end{align}
	generate the conserved total angular-momentum components $J_i=C(\xi_{(i)})$, including both orbital and spin contributions, where $i=1,2,3$ corresponds to $x,y,z$, respectively. We also define the squared total angular momentum as $J^2=J_x^2+J_y^2+J_z^2$ \cite{ZelenkaEtAl2020}. For the metric~\eqref{DW_metric}, the explicit expressions are
	\begin{align}
		E&=-P_t-\frac{A'(r)}{2}S^{tr},
		\label{spinning_energy}\\
		J_x&=-\sin\phi\left(P_\theta+rS^{r\theta}\right)
		-\cos\phi\cot\theta
		\left(P_\phi+r\sin^2\theta\,S^{r\phi}\right)\nonumber\\
		&\quad
		+r^2\sin^2\theta\cos\phi\,S^{\theta\phi},
		\label{spinning_Jx}\\
		J_y&=\cos\phi\left(P_\theta+rS^{r\theta}\right)
		-\sin\phi\cot\theta
		\left(P_\phi+r\sin^2\theta\,S^{r\phi}\right)
		\nonumber\\
		&\quad
		+r^2\sin^2\theta\sin\phi\,S^{\theta\phi},
		\label{spinning_Jy}\\
		J_z&=P_\phi+r\sin^2\theta\,S^{r\phi}
		+r^2\sin\theta\cos\theta\,S^{\theta\phi},
		\label{spinning_Jz}\\
		J^2&=\left(P_\theta+rS^{r\theta}\right)^2
		+\left(\frac{P_\phi}{\sin\theta}
		+r\sin\theta\,S^{r\phi}\right)^2
		+r^4\sin^2\theta\left(S^{\theta\phi}\right)^2,
		\label{spinning_J_squared}
	\end{align}
	where $A'(r)=dA/dr$. The angular-momentum components satisfy the rotational algebra \cite{ZelenkaEtAl2020,DAmbrosiEtAl2015},
	\begin{equation}
		\{J_i,J_j\}=\epsilon_{ijk}J_k.
	\end{equation}
	Thus, the three components $J_i$ cannot all be counted as mutually commuting integrals. Since time translations commute with spatial rotations, $\{E,J_i\}=0$. Together with the rotational algebra, this implies
	\begin{equation}
		\{E,J_z\}=\{E,J^2\}=\{J_z,J^2\}=0.
	\end{equation}
	These integrals can be used to reduce the five-degree-of-freedom covariant system to a system with two degrees of freedom. A detailed example of this reduction in Schwarzschild spacetime can be found in Ref.~\cite{Bizyaev2026}. This reduction does not by itself establish integrability of the remaining dynamics. Since a sufficient number of independent commuting integrals has not been established for the full MPD--TD dynamics in the background considered here, we investigate the possible occurrence of spin-induced chaos using dynamical diagnostics.
    
    %%%%%%%%%%%%%%%%%%%%%%%%%%%%%%%%%%%%%%%%%%%%%%%%%%%%%%%%%%%%%%%%%%%%%%%%%%%%%%%%%%%%%%%%%%%%%%%%%%%%%%%%%%%%%%%%%%%%%%%
    \subsection{Initial conditions}
    \label{sec:initial_conditions}
    %%%%%%%%%%%%%%%%%%%%%%%%%%%%%%%%%%%%%%%%%%%%%%%%%%%%%%%%%%%%%%%%%%%%%%%%%%%%%%%%%%%%%%%%%%%%%%%%%%%%%%%%%%%%%%%%%%%%%%%
    
    To specify the initial conditions, we use spherical symmetry to align the polar axis with the conserved total angular momentum. This coordinate choice does not change the physics of the system. Since the total angular momentum has a fixed direction, a single, time-independent rotation of the coordinates gives
    \begin{equation}
    	J_x=J_y=0,
    	\qquad
    	J_z=J>0.
    	\label{initial_J_alignment}
    \end{equation}
    This condition does not confine the orbit to the equatorial plane. Spin--curvature coupling allows angular momentum to be exchanged between the orbital and spin contributions while their sum remains conserved, so the orbital plane need not remain perpendicular to the polar axis.
    
\begin{figure}[htbp]
    	\centering
    	\includegraphics[width=0.5\linewidth]{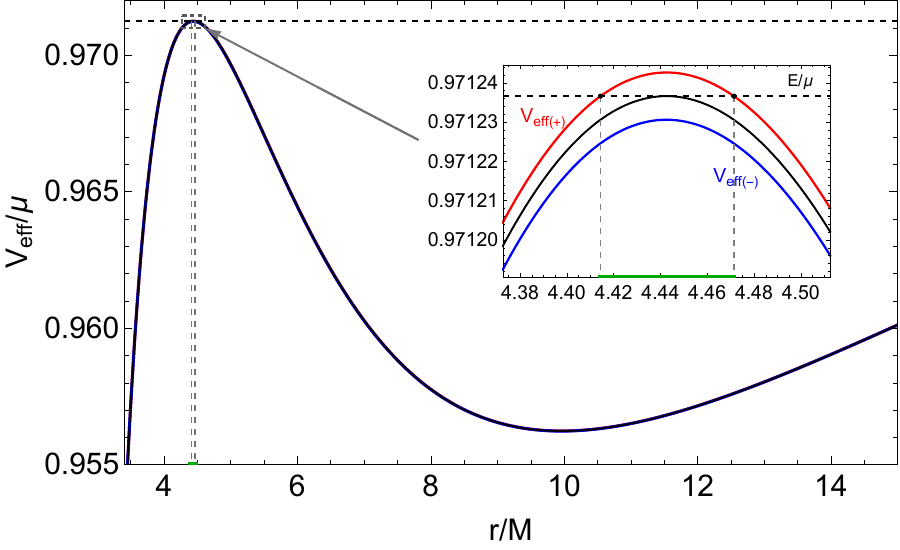}
    	\caption{Effective potentials on the equatorial plane for $\mu=10^{-4}M$, $J=3.8\mu M$, and $\zeta=0.1$. The red and blue solid curves represent $V_{\mathrm{eff}(+)}$ and $V_{\mathrm{eff}(-)}$, respectively, for $s=10^{-4}\mu M$, while the black solid curve shows the nonspinning limit. The black dashed horizontal line indicates the prescribed particle energy $E$. The green segments on the radial axis mark the admissible initial radii at this energy.}
    	\label{fig:effective_potential}
    \end{figure}

    Unlike geodesic motion, the full spinning-particle dynamics cannot generally be described by an ordinary effective potential on the $(r,\theta)$ plane. We can nevertheless obtain bounds on the initial data by setting $P_r=P_\theta=0$ at fixed $(\mu,J,s)$. Following the construction for Schwarzschild spacetime in Refs.~\cite{SuzukiMaeda1997,ZelenkaEtAl2020}, we obtain two branches for the metric~\eqref{DW_metric},
    \begin{subequations}
    	\label{initial_effective_potential}
    	\begin{align}
    		V_{\mathrm{eff}(\pm)}(r,\theta)
    		&=
    		\mu\left[
    		\sqrt{A(r)}\cosh \eta_{(\pm)} 
    		+\frac{rA'(r)\sinh \eta_{(\pm)}}
    		{2\sqrt{A(r)}\cosh \eta_{(\pm)}}
    		\left(
    		\frac{J\sin\theta}{\mu r}
    		-\sinh \eta_{(\pm)}
    		\right)
    		\right],
    		\label{initial_Veff}\\
    		\sinh \eta_{(\pm)}
    		&=
    		\frac{\mu Jr\sin\theta}{\mathcal{P}}
    		\pm
    		\left[
    		\frac{\mu^2J^2r^2\sin^2\theta}{\mathcal{P}^2}
    		-\frac{\mathcal{Q}}{\mathcal{P}}
    		\right]^{1/2},
    		\label{initial_sinh_eta}\\
    		\mathcal{P}
    		&=\mu^2r^2-s^2B(r),\\
    		\mathcal{Q}
    		&=(J^2-s^2)B(r)
    		+[1-B(r)]J^2\sin^2\theta.
    	\end{align}
    \end{subequations}
    Following convention, we call these branches an ``effective potential,'' although they do not govern the full orbital evolution. Figure~\ref{fig:effective_potential} shows the two branches on the equatorial plane. We use them to select initial states with $P_r=0$ and $\theta=\pi/2$. For fixed $(\mu,J,s)$, the two branches bound an allowed band of initial energies at each radius. The boundaries correspond to $P_\theta=0$, while $P_\theta$ is generally nonzero inside the band. For a prescribed energy $E$, the allowed initial radii are the intervals where the horizontal energy line passes through this band. Both branches are therefore needed to determine the available initial data \cite{ZelenkaEtAl2020}. The endpoints of these intervals should not be identified with the pericenter and apocenter of a particular orbit. An allowed initial state near a local maximum of the effective potential may still lead to a plunge, so whether the trajectory remains bound must be checked through numerical evolution.
    
    The nonnegativity of the square-root argument in Eq.~\eqref{initial_sinh_eta} implies
    \begin{equation}
    	|\cos\theta|
    	\leq
    	\frac{s}{J}
    	\left[
    	\frac{\mu^2r^2+J^2-s^2B(r)}
    	{\mu^2r^2+s^2[1-B(r)]}
    	\right]^{1/2},
    	\label{initial_angular_bound}
    \end{equation}
    where we consider the exterior region with $0<B(r)<1$, so that the denominator is positive. To assess this bound for an astrophysical EMRI, we introduce the dimensionless spin $\chi=s/\mu^2$. A Kerr black hole satisfies $\chi\leq1$, while a representative estimate of the maximum spin for a neutron‑star is $\chi\approx0.6$ \cite{Hartl2003}. Thus, in either case, $s\lesssim\mu^2$. For the orbits considered here, $J=\mathcal O(\mu M)$ and $r=\mathcal O(M)$. Since $\mu/M\ll1$, we have
    \begin{equation}
    	\frac{s}{J}\lesssim\mathcal O\left(\frac{\mu}{M}\right)\ll1,
    	\qquad
    	\frac{s}{\mu r}\lesssim\mathcal O\left(\frac{\mu}{M}\right)\ll1.
    \end{equation}
    For $0<B(r)<1$, the above estimates imply that both $s^2B(r)$ and $s^2[1-B(r)]$ are much smaller than $\mu^2r^2$. Neglecting these terms in Eq.~\eqref{initial_angular_bound} gives the leading-order estimate
    \begin{equation}
    	|\cos\theta|
    	\lesssim
    	\frac{s}{J}
    	\sqrt{1+\frac{J^2}{\mu^2r^2}}.
    \end{equation}
    Taken together, these estimates indicate that, for astrophysically realistic spins, the particle's motion is restricted to a narrow region around the equatorial plane. The small ratio $s/J$ characterizes the departure from planar geodesic motion, while the equatorial plane is recovered in the spinless limit.
    
    We study the influence of the NLG deformation parameter $\zeta$ on chaotic motion by selecting initial conditions in the vicinity of the geodesic homoclinic orbit. This orbit forms the separatrix between bound and plunging geodesic motion and is asymptotic to an unstable circular geodesic. In the integrable geodesic limit, the stable and unstable manifolds of this unstable orbit coincide along the homoclinic separatrix. Spin--curvature coupling acts as a perturbation to the geodesic dynamics and can break the coincidence of the stable and unstable manifolds, whose transverse intersections can give rise to chaotic motion \cite{ZelenkaEtAl2020}. The homoclinic neighborhood therefore provides a natural region in which to search for spin-induced chaos.  We fix $J$ and choose the geodesic homoclinic energy \(E_h\) as the reference energy,
    \begin{equation}
    	E_{\mathrm h}=V_{\mathrm{geo}}(r_{\mathrm{un}}).
    \end{equation}
    Here $V_{\mathrm{geo}}$ denotes the spinless limit ($s=0$) of the effective potential in Eq.~\eqref{initial_effective_potential}, and $r_{\mathrm{un}}$ is the radius of its local maximum, corresponding to an unstable circular geodesic. This choice allows us to examine how varying $\zeta$ changes the nearby spinning-particle dynamics, while the actual occurrence of chaos is determined numerically.
    
    Following the initial-condition construction of Ref.~\cite{YuanEtAl2026}, adapted here to the metric~\eqref{DW_metric}, we fix $(\mu,J,s)$ and $E=E_{\mathrm h}$ and choose
    \begin{equation}
    	t_0=\phi_0=0,
    	\qquad
    	\theta_0=\frac{\pi}{2},
    	\qquad
    	P_{r0}=0,
    \end{equation}
    where a subscript $0$ denotes evaluation at $\tau=0$. The initial radius $r_0$ is sampled from the admissible intervals near $r_{\mathrm{un}}$. At this initial position, Eqs.~\eqref{spinning_Jx}--\eqref{spinning_Jz}, with $J_x=J_y=0$ and $J_z=J$, reduce to
    \begin{align}
    	r_0^2S^{\theta\phi}_0&=0,\\
    	P_{\theta0}+r_0S^{r\theta}_0&=0,\\
    	P_{\phi0}+r_0S^{r\phi}_0&=J.
    \end{align}
    Together with the TD condition~\eqref{TD} and $P_{r0}=0$, these relations give
    \begin{equation}
    	\begin{aligned}
    		S^{r\theta}_0&=-\frac{P_{\theta0}}{r_0},
    		\qquad
    		S^{r\phi}_0=\frac{J-P_{\phi0}}{r_0},\\[6pt]
    		S^{tr}_0&=-\frac{P_{\theta0}^2+P_{\phi0}(P_{\phi0}-J)}
    		{r_0P_{t0}},\\[6pt]
    		S^{t\theta}_0&=S^{t\phi}_0=S^{\theta\phi}_0=0.
    	\end{aligned}
    	\label{initial_spin_components}
    \end{equation}
    The initial spin tensor is therefore determined once $(P_{t0},P_{\theta0},P_{\phi0})$ are known. To find these three momentum components, we evaluate the energy integral~\eqref{spinning_energy} and the mass and spin constraints~\eqref{mass_spin} at the initial position:
    \begin{align}
    	0&=E_{\mathrm h}+P_{t0}
    	+\frac{A'(r_0)}{2}S^{tr}_0,
    	\label{initial_energy_constraint}\\
    	0&=\mu-
    	\sqrt{\frac{P_{t0}^2}{A(r_0)}
    		-\frac{P_{\theta0}^2+P_{\phi0}^2}{r_0^2}},
    	\label{initial_mass_constraint}\\
    	0&=s-
    	\sqrt{-\frac{A(r_0)}{B(r_0)}(S^{tr}_0)^2
    		+\frac{r_0^2}{B(r_0)}
    		\left[(S^{r\theta}_0)^2+(S^{r\phi}_0)^2\right]}.
    	\label{initial_spin_constraint}
    \end{align}
    After substituting Eq.~\eqref{initial_spin_components}, Eqs.~\eqref{initial_energy_constraint}--\eqref{initial_spin_constraint} form three coupled algebraic equations for $(P_{t0},P_{\theta0},P_{\phi0})$. We solve them simultaneously using Newton's method with an analytical Jacobian and at least 128-bit floating-point precision. This determines the initial values of all 14 evolved components: four coordinates $x^a$, four momentum components $P^a$, and six independent components of the antisymmetric spin tensor $S^{ab}$.
    
    \begin{figure}[ht]
    	\centering
    	\begin{tabular}{@{}c@{\hspace{0.02\textwidth}}c@{}}
    		\includegraphics[width=0.4\textwidth]{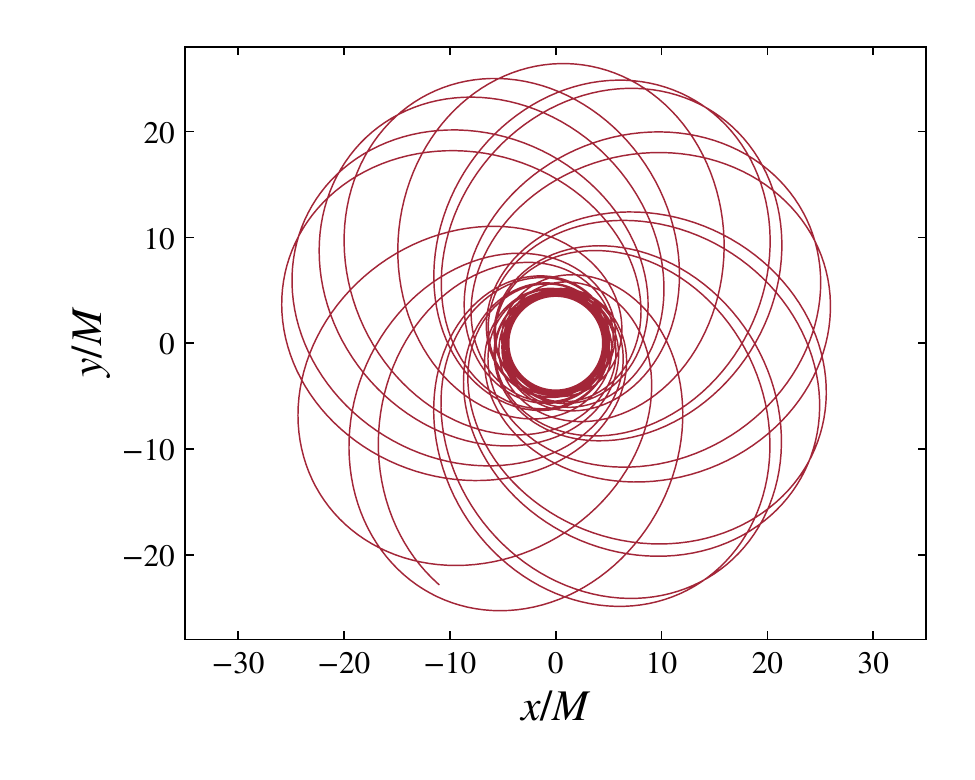}
    		&
    		\includegraphics[width=0.4\textwidth]{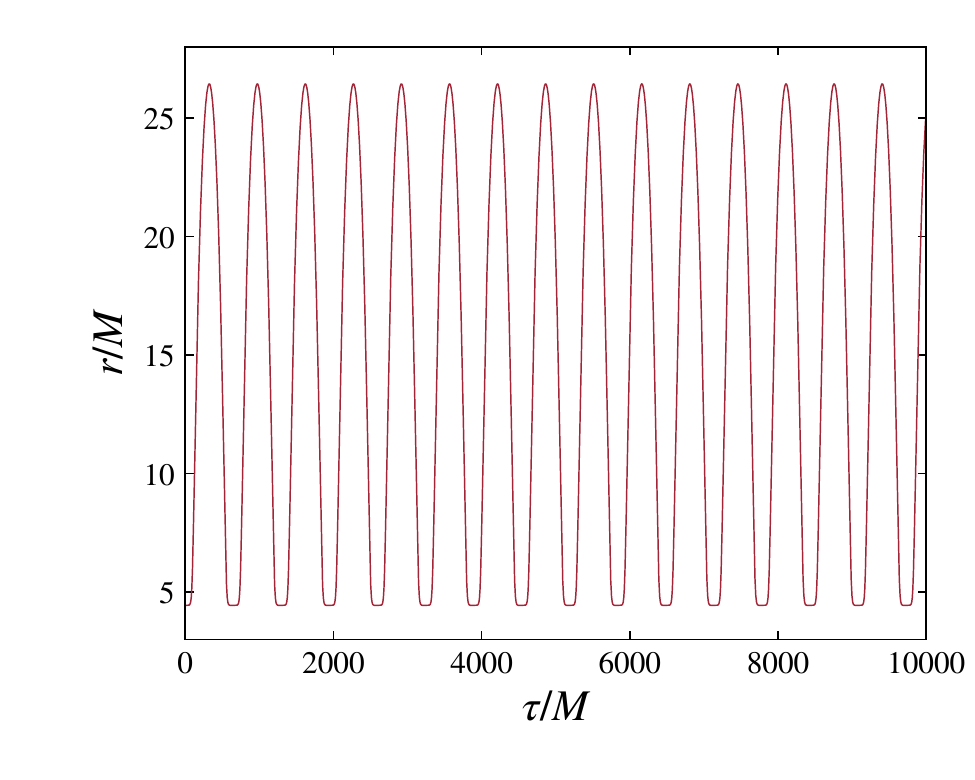}
    		\\[-2mm]
    		(a) & (b)
    		\\[2mm]
    		\includegraphics[width=0.4\textwidth]{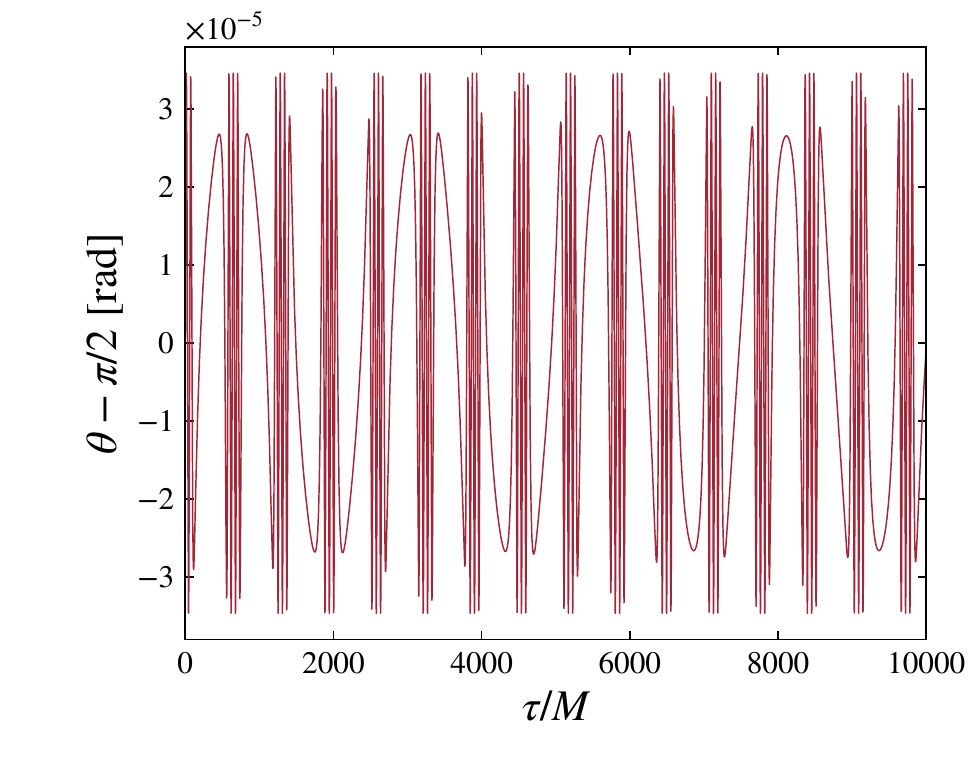}
    		&
    		\includegraphics[width=0.4\textwidth]{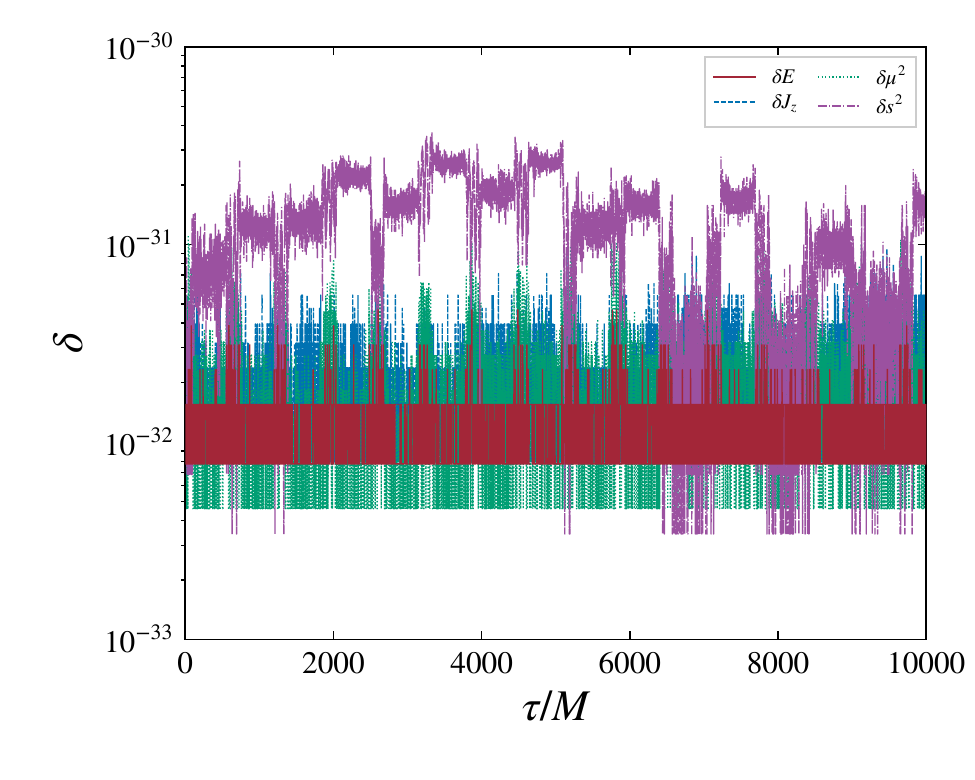}
    		\\[-2mm]
    		(c) & (d)
    	\end{tabular}
    	
    	\caption{Representative orbit and conservation errors for $n=2$, $\zeta=0.1$, $\mu=10^{-4}M$, $J_z=3.8\mu M$, and $s=10^{-4}\mu M$. The initial conditions are $r_0=4.442628672M$, $\theta_0=\pi/2$, $t_0=\phi_0=0$, and $P_{r0}=0$, with $E=E_{\mathrm h}\simeq0.971236576\mu$. The integration extends to $\tau=10^4M$ with $\Delta\tau=M$. Panels show (a) the orbital projection, (b) radial evolution, (c) deviation from the equatorial plane, and (d) relative conservation errors evaluated at integration step points.}
    	\label{fig:representative_orbit}
    \end{figure}
    
    We integrate the MPD equations in Julia. The initial constraint equations are solved with \texttt{NLsolve.jl}, and the subsequent evolution is computed with \texttt{DifferentialEquations.jl} using the sixteenth-order implicit Gauss--Legendre Runge--Kutta solver \texttt{IRKGL16} from \texttt{IRKGaussLegendre.jl}. Equations~\eqref{MPD1}--\eqref{MPD2} are evolved together with $dx^a/d\tau=u^a$, with the four-velocity evaluated from Eq.~\eqref{velocity_momentum}. We use \texttt{Double64} arithmetic from \texttt{DoubleFloats.jl}, providing approximately 32 decimal digits of precision, and a fixed proper-time step $\Delta\tau=1$ in units of $M=1$. Numerical conservation is checked at the integration step points by comparing $E$, $J_z$, $\mu^2$, and $s^2$, defined in Eqs.~\eqref{spinning_energy}, \eqref{spinning_Jz}, and \eqref{mass_spin}, with their prescribed values.
    
    A representative trajectory is shown in Fig.~\ref{fig:representative_orbit}. The particle repeatedly circles near pericenter before moving outward, producing the precessing orbit seen in the projection. Its radial coordinate varies between approximately $4.443M$ and $26.437M$, with no plunge during the integration interval. The motion stays close to the equatorial plane, with $|\theta-\pi/2|\lesssim3.5\times10^{-5}$. All four relative conservation errors remain below $10^{-30}$. Whether this trajectory is regular or chaotic cannot be determined from these plots alone; we address this question using the diagnostics introduced below.
	
	%%%%%%%%%%%%%%%%%%%%%%%%%%%%%%%%%%%%%%%%%%%%%%%%%%%%%%%%%%%%%%%%%%%%%%%%%%%%%%%%%%%%%%%%%%%%%%%%%%%%%%%%%%%%%%%%%%%%%%%
	\section{Dynamical diagnostics}
	\label{sec:diagnostics}
	%%%%%%%%%%%%%%%%%%%%%%%%%%%%%%%%%%%%%%%%%%%%%%%%%%%%%%%%%%%%%%%%%%%%%%%%%%%%%%%%%%%%%%%%%%%%%%%%%%%%%%%%%%%%%%%%%%%%%%%
	
	%%%%%%%%%%%%%%%%%%%%%%%%%%%%%%%%%%%%%%%%%%%%%%%%%%%%%%%%%%%%%%%%%%%%%%%%%%%%%%%%%%%%%%%%%%%%%%%%%%%%%%%%%%%%%%%%%%%%%%%
	\subsection{Poincaré sections and rotation numbers}
	\label{sec:poincare_rotation}
	%%%%%%%%%%%%%%%%%%%%%%%%%%%%%%%%%%%%%%%%%%%%%%%%%%%%%%%%%%%%%%%%%%%%%%%%%%%%%%%%%%%%%%%%%%%%%%%%%%%%%%%%%%%%%%%%%%%%%%%
	As discussed in Sec.~\ref{sec:integrability}, the dynamics can be reduced to a system with two degrees of freedom. For this reduced system, conservation of the Hamiltonian confines the motion to a three‑dimensional hypersurface in the four‑dimensional phase space. Successive intersections with a transverse two‑dimensional surface define a Poincaré map. We take the equatorial plane as the section and retain crossings with positive polar momentum,
	\begin{equation}
		\theta=\frac{\pi}{2},
		\qquad
		P_\theta>0.
		\label{poincare_section}
	\end{equation}
	The successive intersections are plotted in the $(r,P_r)$ plane. To interpret their distribution, we first consider the integrable limit, where bounded motion lies on an invariant torus. This torus can be pictured as a doughnut‑shaped surface in phase space, with motion around its two independent angular directions described by the fundamental frequencies $\omega_1$ and $\omega_2$. Their ratio,
	\begin{equation}
		\omega=\frac{\omega_1}{\omega_2},
	\end{equation}
	determines whether the motion is quasiperiodic or periodic. If $\omega$ is irrational, the motion never repeats exactly. A single trajectory densely covers the torus, and its section points trace a closed invariant curve. The motion is then quasiperiodic, and the torus is called nonresonant. If $\omega$ is rational, the two angular motions repeat after a common period. The torus is resonant, and an individual periodic orbit gives a finite set of points on the section.
	
	These two types of tori respond differently to a weak perturbation. For a sufficiently small, sufficiently smooth Hamiltonian perturbation of an integrable system, the Kolmogorov--Arnold--Moser (KAM) theorem guarantees the persistence of many invariant tori with Diophantine frequency ratios \cite{Wiggins2003}. Their section points still lie on closed curves. Under a weak perturbation, resonant tori can break up, leaving periodic orbits as required by the Poincar\'e--Birkhoff theorem. In the generic nondegenerate case, these form chains of alternating elliptic and hyperbolic periodic points. Near an elliptic point, regular trajectories trace nested closed curves that form a stability island. Under forward evolution, trajectories on the stable manifold approach the hyperbolic point, while those on the unstable manifold move away from it locally. Transverse homoclinic intersections of the stable and unstable manifolds of a hyperbolic periodic orbit imply the existence of chaotic dynamics. The section points of a chaotic trajectory no longer lie on a smooth invariant curve, but spread over a region of the section. Surviving invariant curves, regular resonant islands, and chaotic layers can therefore appear together on the same section. Further resonances may produce smaller island chains, giving a finer mixture of regular and chaotic regions.
	
	As a complementary tool, the rotation number provides a quantitative way to locate resonances and distinguish regular from chaotic motion. Let $Q_i=(r_i,P_{r,i})$ denote successive section points and $Q_{\mathrm c}$ the central elliptic fixed point. We define the displacement vectors \(R_i=Q_i-Q_{\mathrm c}\), and measure the angular advance $\vartheta_i$ from $R_i$ to $R_{i+1}$ using a consistent orientation. The rotation number is defined as
	\begin{equation}
		\nu=\lim_{N\to\infty}\nu_N,
		\qquad
		\nu_N=\frac{1}{2\pi N}\sum_{i=1}^{N}\vartheta_i.
		\label{rotation_number}
	\end{equation}
	For motion on a regular invariant torus, $\nu$ equals the frequency ratio $\omega$ modulo an integer \cite{ZelenkaEtAl2020}. We obtain the rotation curve by plotting $\nu_N$ against the initial radius. Smooth segments indicate regular motion, while plateaus at rational values identify resonant islands. In chaotic layers, the estimated rotation number can vary irregularly between nearby initial conditions.
	
	\begin{figure}[ht]
		\centering
		\includegraphics[width=\textwidth]
		{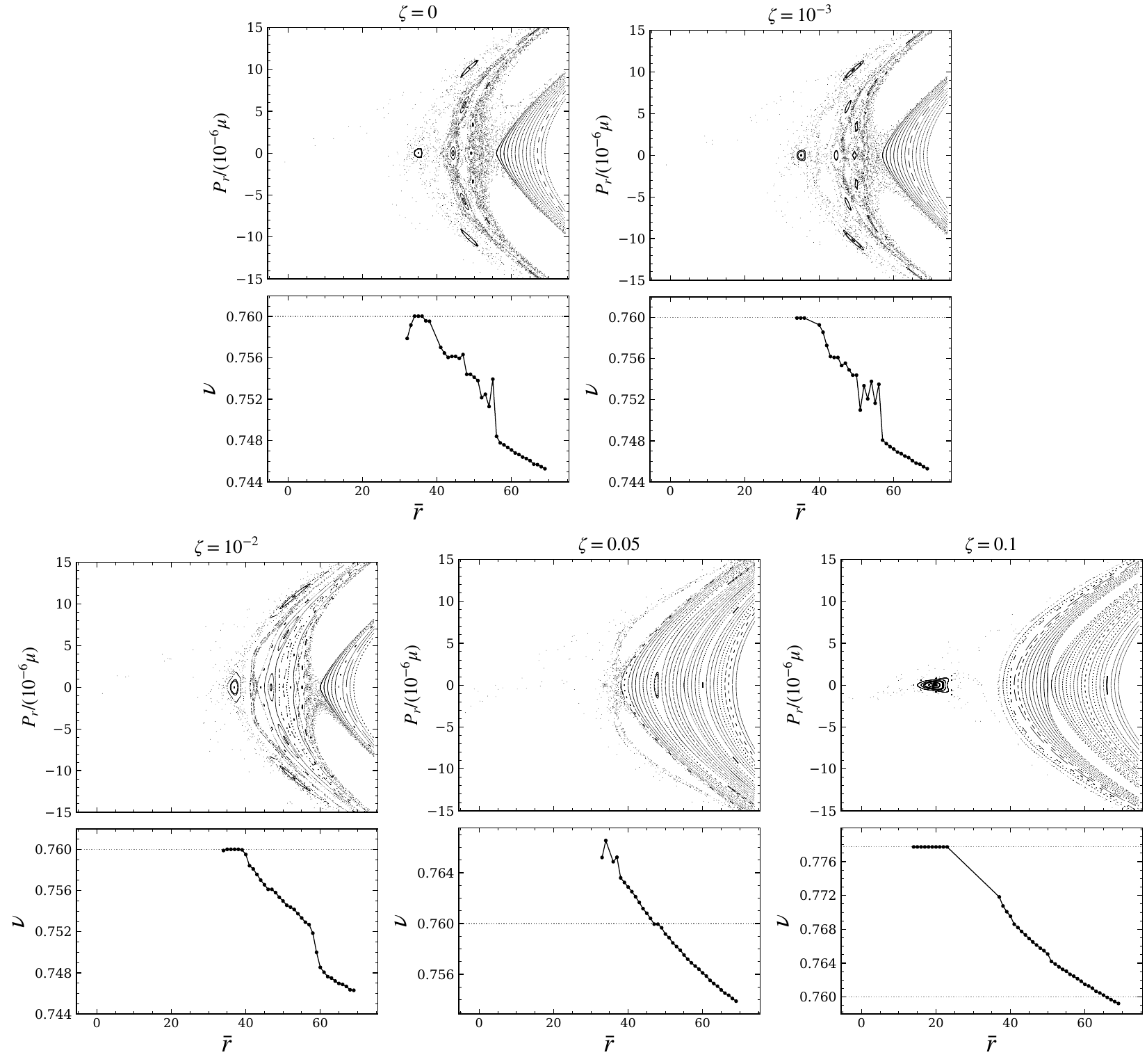}
		\caption{Poincar\'e sections and corresponding rotation curves for $J_z=3.8\mu M$, $s=10^{-4}\mu M$, and $E=E_{\mathrm h}(\zeta)$. All trajectories are integrated up to $\tau=10^6M$ or until plunge. Rotation numbers are shown only for trajectories with at least 600 section crossings. Dotted horizontal lines mark $\nu=19/25$ and, for $\zeta=0.1$, additionally $\nu=7/9$.}
		\label{fig:poincare-rotation-four-zeta}
	\end{figure}
	
	We set $E=E_{\mathrm h}(\zeta)$, the energy at the maximum of the spinless effective potential, to compare trajectories near the associated homoclinic orbit for each spacetime background. The height of this potential barrier depends on $\zeta$; fixing a common numerical energy would place orbits at different relative positions with respect to the barrier. The radial position of the potential peak also shifts with $\zeta$, so direct comparison in the original $r$ coordinate carries this global offset. We adopt the shifted coordinate
	\begin{equation}
		\bar r=\frac{r-r_{\rm ref}(\zeta)}{10^{-6}M},
		\label{bar_r}
	\end{equation}
	where $r_{\rm ref}(\zeta)$ is the radial location of the spinless effective-potential maximum, rounded to six decimal places. Starting from \(r_{\rm ref}(\zeta)\), we sample 70 initial radii with step \(\Delta r_0=10^{-6}M\), giving \(\bar r_0=0,\,1,\,\ldots,\,69\). This coordinate measures radial offsets from the starting point. It is also applied to both the section crossings and the initial radii, allowing their locations in the Poincar\'e sections and rotation curves to be compared directly in Figure~\ref{fig:poincare-rotation-four-zeta}. We inspect this figure and identify distinct characteristics for different values of the deformation parameter.
	
	For $\zeta=0$ (GR), smooth invariant curves occupy the outer part of the Poincar\'e section, while resonant islands and scattered points coexist in the inner region. At $\zeta=10^{-3}$, the section and rotation curve remain similar to those in GR. Changes in the scattered region and the resonant structures are difficult to see at the scale shown. At $\zeta=10^{-2}$, the scattered region becomes smaller, regular curves are easier to see, and the rotation curve fluctuates less, while the plateau near $\nu=19/25$ remains. At $\zeta=0.05$, the rotation curve is mostly smooth outside its inner edge. The plateau near $\nu=19/25$ is narrow and difficult to see with the current sampling, but the corresponding resonant island structure is still visible in the section. At $\zeta=0.1$, nested closed curves form a stability island in the inner region, corresponding to a wider plateau near $\nu=7/9$.
	
	The overall structure changes little at $\zeta=10^{-3}$. From $\zeta=10^{-2}$ onward, the smaller scattered region and smoother rotation curves suggest less irregular motion within the sampled region. However, these features alone are not enough to classify the orbits as regular or chaotic. We therefore use the fast Lyapunov indicator (FLI) to check whether these changes are accompanied by less detectable chaos over the finite integration interval.

    %%%%%%%%%%%%%%%%%%%%%%%%%%%%%%%%%%%%%%%%%%%%%%%%%%%%%%%%%%%%%%%%%%%%%%%%%%%%%%%%%%%%%%%%%%%%%%%%%%%%%%%%%%%%%%%%%%%%%%%
	\subsection{Fast Lyapunov indicator}
	\label{sec:FLI}
	%%%%%%%%%%%%%%%%%%%%%%%%%%%%%%%%%%%%%%%%%%%%%%%%%%%%%%%%%%%%%%%%%%%%%%%%%%%%%%%%%%%%%%%%%%%%%%%%%%%%%%%%%%%%%%%%%%%%%%%
	The maximal Lyapunov exponent measures the asymptotic exponential growth rate of an infinitesimal perturbation. In the variational method, the variational equations are integrated together with the equations of motion to follow the perturbation along a reference trajectory \cite{WuHuang2003,WuEtAl2006}. Obtaining a converged exponent, however, often requires long integrations, making extensive scans over initial conditions computationally demanding. We therefore use the fast Lyapunov indicator (FLI), which records the accumulated logarithmic growth of the orbital separation without dividing by the elapsed time. It can distinguish regular and chaotic motion without waiting for the Lyapunov exponent to converge, making it useful for such scans \cite{WuEtAl2006,Han2008}. 
	
	For relativistic orbits, however, coordinate-dependent definitions of orbital separation are problematic and may obscure chaotic behavior \cite{WuHuang2003,Han2008}. A relativistic two-trajectory method addresses this issue by measuring spatial separation in the reference observer's local rest frame and expressing its growth in terms of that observer's proper time \cite{WuHuang2003}. This approach was subsequently extended to the FLI \cite{WuEtAl2006} and applied to spinning-particle dynamics \cite{Han2008,YuanEtAl2026}. Rather than integrating variational equations to follow an infinitesimal perturbation, the two-trajectory method approximates its evolution by tracking the difference between nearby trajectories. We adopt this method to calculate the FLI using the same numerical solver as for the orbital evolution.
	
	In our implementation, the reference trajectory is initialized and evolved following the procedure described in Sec.~\ref{sec:initial_conditions}. A neighbouring trajectory is constructed by shifting the initial radius by $\delta r(0)=10^{-16}M$ and solving the constraints again with the same prescribed conserved quantities, then evolved using the same numerical scheme.  At equal values of the proper-time parameter $\tau$, the coordinate displacement between the neighbouring and reference trajectories is
	\begin{equation}
		\Delta x^a(\tau)=\widetilde{x}^a(\tau)-x^a(\tau),
	\end{equation}
	where a tilde denotes the neighbouring trajectory. To measure the spatial separation in the reference particle's local rest frame, we introduce the projection operator
	\begin{equation}
		h^a{}_b=\delta^a{}_b+u^a u_b,
		\qquad
		h_{ab}=g_{ab}+u_a u_b.
	\end{equation}
	The projected distance is
	\begin{equation}
		\ell(\tau)=\sqrt{\left|h_{ab}\Delta x^a\Delta x^b\right|},
		\label{FLI_distance}
	\end{equation}
	where the metric and normalized four-velocity are evaluated along the reference trajectory. Within this two-trajectory framework, the FLI is defined as
	\begin{equation}
		\mathrm{FLI}(\tau)=\log_{10}\frac{\ell(\tau)}{\ell(0)}.
		\label{FLI_definition}
	\end{equation}
	From Eq.~\eqref{FLI_definition}, exponential growth of the separation gives approximately linear growth of the FLI, whereas algebraic separation gives logarithmic growth \cite{WuEtAl2006}. The FLI can therefore be used to distinguish these growth patterns without requiring a converged estimate of the Lyapunov exponent.
	
	The distance in Eq.~\eqref{FLI_distance} is a local measure: the coordinate difference approximates a connecting displacement only while the trajectories remain sufficiently close. As the separation grows, this approximation eventually breaks down. For motion confined to a finite region, the separation also eventually saturates and no longer reflects the local exponential divergence. We therefore rescale the state difference when the distance first exceeds $\ell_{\max}=10^{-10}M$ at an integration step point.
	
	For this purpose, we denote the full states of the reference and neighbouring trajectories by $\mathcal Z$ and $\widetilde{\mathcal Z}$, respectively. Each contains four coordinates, four momentum components, and six independent spin-tensor components. If $\ell_j$ is the separation immediately before the $j$th rescaling, we update the neighbouring state as
	\begin{equation}
		\widetilde{\mathcal Z}
		\longleftarrow
		\mathcal Z
		+\frac{\ell(0)}{\ell_j}
		\left(\widetilde{\mathcal Z}-\mathcal Z\right).
		\label{FLI_rescaling}
	\end{equation}
	The reference state is unchanged. All components of the state difference are multiplied by the same positive factor, preserving its direction. This linear rescaling does not explicitly enforce the TD condition~\eqref{TD} or the prescribed mass and spin magnitude~\eqref{mass_spin}, so the updated neighbouring state need not satisfy these constraints exactly. It is therefore not an exact projection onto the nonlinear constraint surface \cite{YuanEtAl2026}. The resulting constraint violation is taken to be negligible for the small separations used here. The growth removed by each rescaling is retained in the accumulated indicator. After k rescalings,
	\begin{equation}
		\mathrm{FLI}(\tau)
		=\sum_{j=1}^{k}\log_{10}\frac{\ell_j}{\ell(0)}
		+\log_{10}\frac{\ell(\tau)}{\ell(0)},
		\label{FLI_accumulated}
	\end{equation}
    where the sum accounts for the growth removed by previous resets and the final term measures the growth since the latest reset. For trajectories reaching $\tau=10^6M$, we adopt $\mathrm{FLI}>15$ as the finite-time criterion for chaotic motion. Plunging trajectories are treated separately.
    
    \begin{figure}[ht]
    	\centering
    	\includegraphics[width=0.6\linewidth]{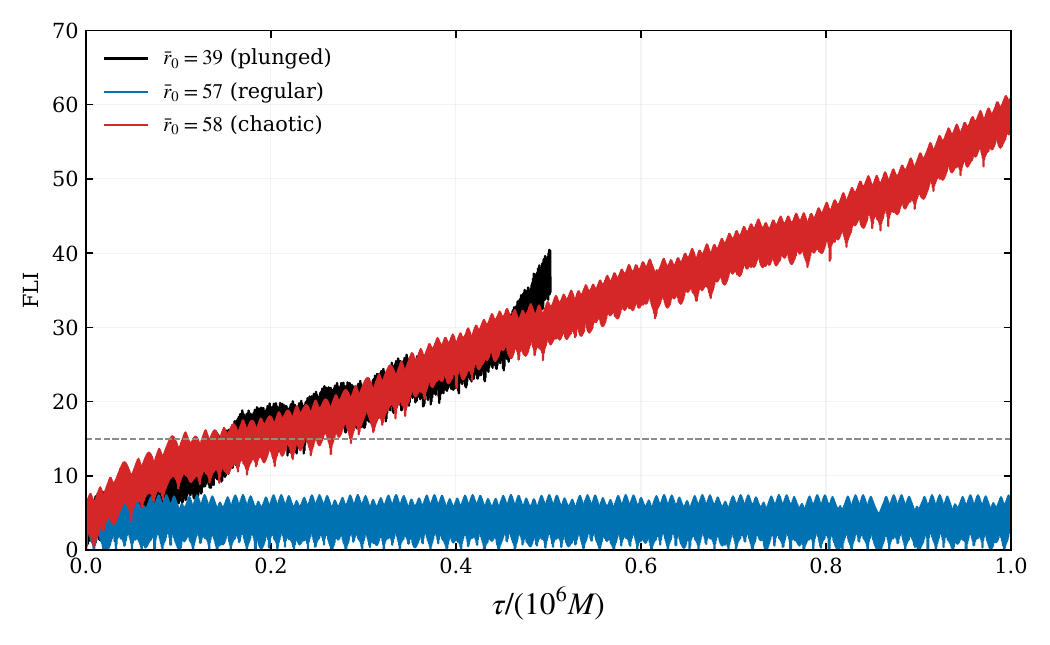}
    	\caption{Time evolution of the FLI for three representative trajectories at $\zeta=10^{-2}$, with $J_z=3.8\mu M$, $s=10^{-4}\mu M$, and $E=E_{\mathrm h}(0.01)$. The initial radial conditions are $\bar r_0=39$ (black), $57$ (blue), and $58$ (red). The black curve terminates at $\tau\simeq5\times10^5M$ when the plunge criterion is triggered, while the other two trajectories are integrated to $\tau=10^6M$. The horizontal dashed line marks the adopted finite-time threshold $\mathrm{FLI}=15$.}
    	\label{fig:fli-time-zeta001}
    \end{figure}
    
    Figure~\ref{fig:fli-time-zeta001} shows FLI time‑evolution curves for three trajectories taken from the $\zeta=10^{-2}$ parameter scan. Despite having neighbouring initial conditions, $\bar r_0=57$ and 58 display strikingly different dynamical behaviour. For $\bar r_0=57$, the FLI stays low throughout the integration, oscillating but showing no steady increase. This behaviour is characteristic of regular motion. For $\bar r_0=58$, oscillations are superimposed on a steady rising trend, and the FLI reaches roughly 61 by $\tau=10^6M$. This persistent growth marks the orbit as chaotic. The $\bar r_0=39$ orbit also develops considerable FLI growth, hitting about $36.7$ before the plunge condition halts the numerical integration. The regular trajectory experiences a single renormalization at $\tau=876M$, while the chaotic and plunging orbits undergo ten and six renormalizations, respectively. Comparing these cases, chaotic trajectories generally trigger more renormalizations. Nevertheless, the number of renormalizations alone cannot serve as a chaos discriminant.
       
    \begin{figure}[htbp]
    	\centering
    	\includegraphics[width=\textwidth]{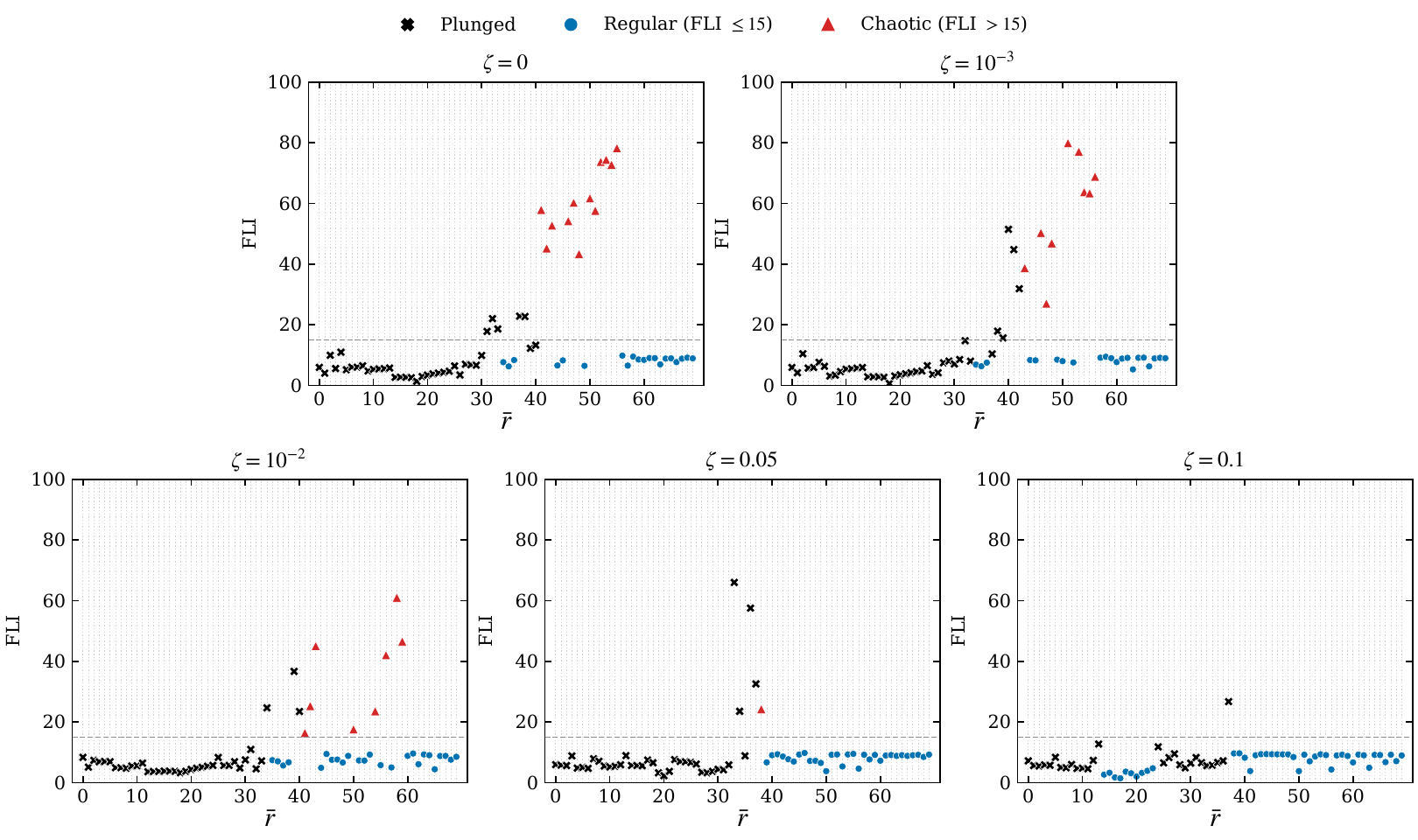}
    	\caption{FLI values for $J_z=3.8\mu M$, $s=10^{-4}\mu M$, and $E=E_{\mathrm h}(\zeta)$. Non-plunging trajectories are integrated to $\tau=10^6M$, while plunging trajectories are recorded before termination. The horizontal dashed line marks the finite-time classification threshold $\mathrm{FLI}=15$.
    	}
    	\label{fig:fli-four-zeta}
    \end{figure}      
    
    Building on these representative single‑orbit examples, we further perform FLI scans across broad ranges of initial radii for several choices of $\zeta$, as illustrated in Figure~\ref{fig:fli-four-zeta}. These FLI‑based results back the distinction between regular and chaotic regions inferred from the Poincar\'e sections and rotation curves, for the same trajectories displayed in Figure~\ref{fig:poincare-rotation-four-zeta}. When \(\zeta=0\) and \(10^{-3}\), high‑FLI trajectories occur among low‑FLI trajectories in the inner region, consistent with the mixture of chaotic layers and stability islands. The outer invariant curves, by contrast, correspond to low FLI values for regular trajectories. At \(\zeta=10^{-2}\), chaotic trajectories remain present, although their locations within the sampled interval change and these chaotic orbits generally have lower FLI values. Only one chaotic trajectory remains at $\zeta=0.05$, while no non-plunging trajectory exceeds the FLI threshold at $\zeta=0.1$. Among the non-plunging trajectories, the fraction classified as chaotic decreases from $37.5\%$ in GR ($\zeta=0$) to $30.0\%$, $24.2\%$, $3.1\%$, and $0\%$ as $\zeta$ increases. These results indicate a trend toward weaker detectable finite-time chaos within the chosen parameter range and initial-condition family.
    
    %%%%%%%%%%%%%%%%%%%%%%%%%%%%%%%%%%%%%%%%%%%%%%%%%%%%%%%%%%%%%%%%%%%%%%%%%%%%%%%%%%%%%%%%%%%%%%%%%%%%%%%%%%%%%%%%%%%%%%%
    \section{Gravitational waveform signatures}
    \label{sec:waveforms}
    %%%%%%%%%%%%%%%%%%%%%%%%%%%%%%%%%%%%%%%%%%%%%%%%%%%%%%%%%%%%%%%%%%%%%%%%%%%%%%%%%%%%%%%%%%%%%%%%%%%%%%%%%%%%%%%%%%%%%%%
    
    %%%%%%%%%%%%%%%%%%%%%%%%%%%%%%%%%%%%%%%%%%%%%%%%%%%%%%%%%%%%%%%%%%%%%%%%%%%%%%%%%%%%%%%%%%%%%%%%%%%%%%%%%%%%%%%%%%%%%%%
    \subsection{Waveform construction and spectral features}
    \label{sec:waveform_construction}
    %%%%%%%%%%%%%%%%%%%%%%%%%%%%%%%%%%%%%%%%%%%%%%%%%%%%%%%%%%%%%%%%%%%%%%%%%%%%%%%%%%%%%%%%%%%%%%%%%%%%%%%%%%%%%%%%%%%%%%%
    
    The orbital analysis above shows that background deformation changes the resonant structures and the occurrence of finite-time chaos within the sampled initial-condition family. We now ask whether these dynamical changes have a corresponding signature in the approximate gravitational radiation. In particular, we examine whether neighboring trajectories with different FLI classifications also differ in their spectral structure, and how this contrast changes when the background deformation alters their classifications. The FLI provides an independent identification of the orbital dynamics, while the waveform analysis tests how those dynamics are reflected in the radiation. We construct approximate gravitational waveforms from numerical MPD trajectories using the mass-quadrupole numerical kludge prescription \cite{BabakEtAl2007}. Orbital coordinates are mapped to an auxiliary flat-space Cartesian system,
    \begin{equation}
    	x_{\mathrm p}=r\sin\theta\cos\phi,
    	\qquad
    	y_{\mathrm p}=r\sin\theta\sin\phi,
    	\qquad
    	z_{\mathrm p}=r\cos\theta,
    	\label{GW_Cartesian}
    \end{equation}
    where the subscript $\mathrm p$ labels the particle. In this section, we denote the particle’s Cartesian position components by $x_{\mathrm p}^i$, where $i,j=1,2,3$, such that
    \begin{equation}
    	x_{\mathrm p}^1=x_{\mathrm p},
    	\qquad
    	x_{\mathrm p}^2=y_{\mathrm p},
    	\qquad
    	x_{\mathrm p}^3=z_{\mathrm p}.
    \end{equation}
    The orbital mass quadrupole is
    \begin{equation}
    	I_{ij}(t_{\mathrm p})
    	=\mu x_{\mathrm p\,i}(t_{\mathrm p})
    	x_{\mathrm p\,j}(t_{\mathrm p}),
    	\label{GW_quadrupole}
    \end{equation}
    where $t_{\mathrm p}$ denotes the coordinate time along the
    particle trajectory. In the far-zone approximation, the Cartesian
    waveform components at observer time $t$ are
    \begin{equation}
    	\mathcal H_{ij}(t)
    	=\frac{2}{D}
    	\left.
    	\frac{d^2 I_{ij}}{dt_{\mathrm p}^2}
    	\right|_{t_{\mathrm p}=t-D}
    	=\frac{2\mu}{D}
    	\left[
    	\ddot x_{\mathrm p\,i}x_{\mathrm p\,j}
    	+2\dot x_{\mathrm p\,i}\dot x_{\mathrm p\,j}
    	+x_{\mathrm p\,i}\ddot x_{\mathrm p\,j}
    	\right]_{t_{\mathrm p}=t-D},
    	\label{GW_Cartesian_components}
    \end{equation}
    where overdots denote differentiation with respect to
    $t_{\mathrm p}$, and $G=c=1$ is assumed.
    Since our numerical trajectories are parametrized by proper time
    $\tau$, we evaluate these derivatives using
    \begin{equation}
    	\frac{d^2}{dt_{\mathrm p}^2}
    	=\frac{1}{(dt_{\mathrm p}/d\tau)^2}
    	\frac{d^2}{d\tau^2}
    	-\frac{d^2t_{\mathrm p}/d\tau^2}
    	{(dt_{\mathrm p}/d\tau)^3}
    	\frac{d}{d\tau}.
    	\label{GW_time_conversion}
    \end{equation}
    For an observer located at $(\Theta,\Phi)$, we adopt the orthonormal triad
    \begin{align}
    	\boldsymbol e_N
    	&=(\sin\Theta\cos\Phi,\,
    	\sin\Theta\sin\Phi,\,
    	\cos\Theta),\\
    	\boldsymbol e_\Theta
    	&=(\cos\Theta\cos\Phi,\,
    	\cos\Theta\sin\Phi,\,
    	-\sin\Theta),\\
    	\boldsymbol e_\Phi
    	&=(-\sin\Phi,\,\cos\Phi,\,0).
    \end{align}
    Here $\boldsymbol e_N$ points along the line of sight to the observer, and $\boldsymbol e_\Theta,\boldsymbol e_\Phi$ span the transverse plane. The plus and cross polarization tensors are defined as
    \begin{equation}
    	\epsilon_+^{ij}
    	=e_\Theta^i e_\Theta^j-e_\Phi^i e_\Phi^j,
    	\qquad
    	\epsilon_\times^{ij}
    	=e_\Theta^i e_\Phi^j+e_\Phi^i e_\Theta^j.
    \end{equation}
    Both tensors are transverse to $\boldsymbol e_N$ and trace-free. The two polarization amplitudes are obtained by contracting $\mathcal H_{ij}$ with these tensors:
    \begin{equation}
    	h_+=\frac12\epsilon_+^{ij}\mathcal H_{ij},
    	\qquad
    	h_\times=\frac12\epsilon_\times^{ij}\mathcal H_{ij}.
    	\label{GW_polarization_projection}
    \end{equation}
    This contraction isolates the transverse-traceless part of $\mathcal H_{ij}$. Written out explicitly,
    \begin{align}
    	h_+
    	&=
    	\frac{1+\cos^2\Theta}{4}
    	\left[
    	(\mathcal H_{xx}-\mathcal H_{yy})\cos2\Phi
    	+2\mathcal H_{xy}\sin2\Phi
    	\right]
    	\nonumber\\
    	&\quad
    	-\frac{\sin^2\Theta}{4}
    	\left(\mathcal H_{xx}+\mathcal H_{yy}-2\mathcal H_{zz}\right)
    	-\sin\Theta\cos\Theta
    	\left(\mathcal H_{xz}\cos\Phi+\mathcal H_{yz}\sin\Phi\right),\\
    	h_\times
    	&=
    	\cos\Theta
    	\left[
    	-\frac12(\mathcal H_{xx}-\mathcal H_{yy})\sin2\Phi
    	+\mathcal H_{xy}\cos2\Phi
    	\right]
    	\nonumber\\
    	&\quad
    	+\sin\Theta
    	\left(\mathcal H_{xz}\sin\Phi-\mathcal H_{yz}\cos\Phi\right).
    	\label{GW_polarizations}
    \end{align}
    Having specified the waveform construction, we now apply it to a comparison of two neighboring trajectories. We select the initial-condition labels $\bar r_0=54$ and $57$. The former is classified as chaotic for $\zeta=0$, $10^{-3}$, and $10^{-2}$, and as regular for $\zeta=0.05$ and $0.1$, whereas the latter remains regular in all five backgrounds. For each fixed value of $\zeta$, the two trajectories share the same particle parameters, energy, and observer configuration. Their comparison therefore controls for these quantities and provides a within-background test of how differences in dynamical classification are reflected in the radiation spectra. We repeat this comparison for all five backgrounds and examine how the spectral contrast of the pair changes when the classification of the $\bar r_0=54$ trajectory changes from chaotic to regular.
    
\begin{figure}[htbp]
    	\centering
    	\includegraphics[width=\textwidth]{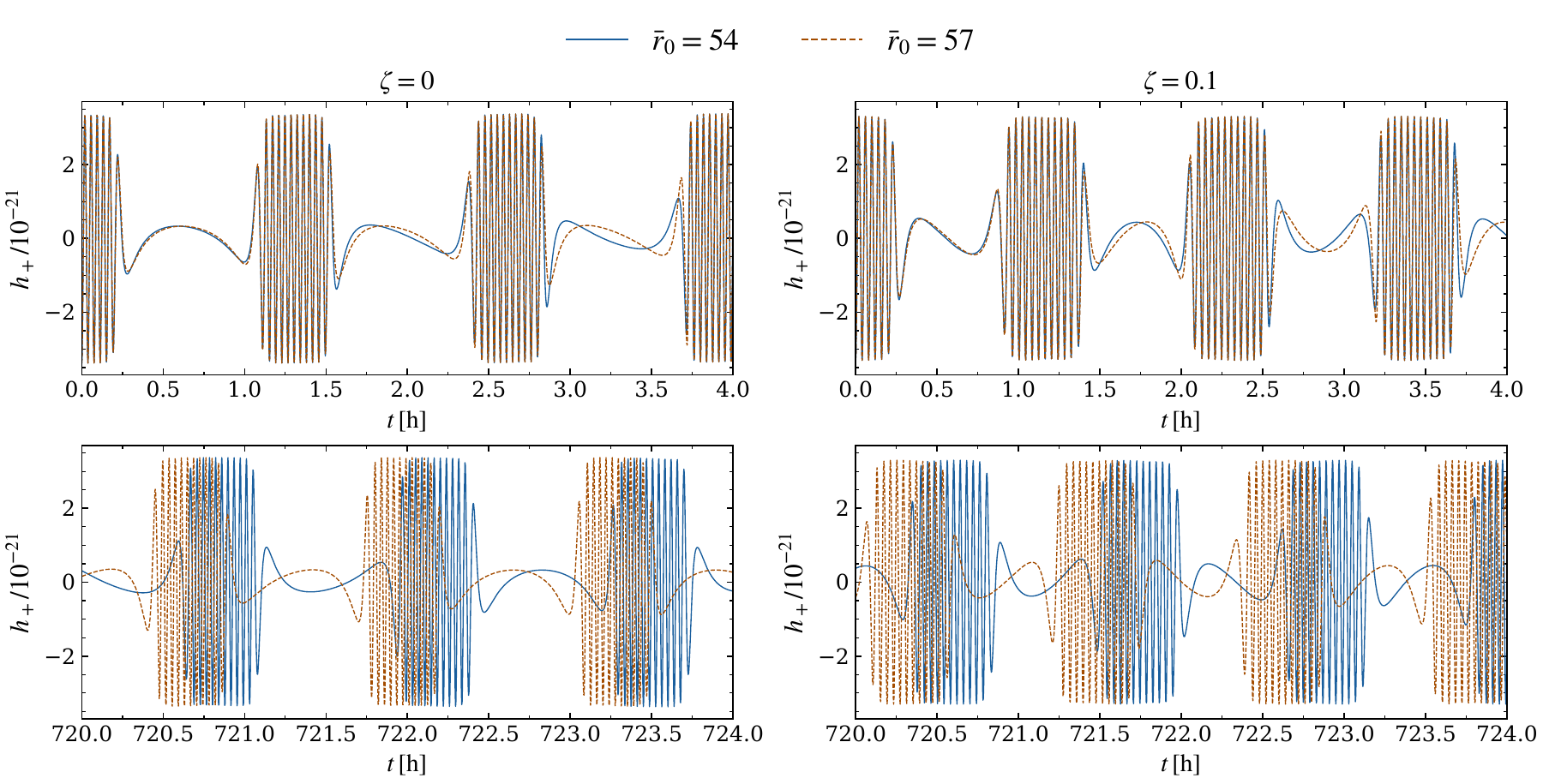}
    	\caption{Waveforms of $h_+$ for the trajectories with $\bar r_0=54$ and $57$ in GR (left) and at $\zeta=0.1$ (right). The upper panels show the first four hours, and the lower panels show a four-hour interval beginning after 30 days. The observer direction is $(\Theta,\Phi)=(\pi/4,0)$. The central mass, mass ratio, and source distance are $M=10^6M_\odot$, $\mu/M=10^{-4}$, and $D=1\,\mathrm{Gpc}$, respectively.}
    	\label{fig:waveforms-gr-zeta01}
    \end{figure}

    As shown in Figure~\ref{fig:waveforms-gr-zeta01}, the waveforms for $\bar r_0=54$ and $57$ appear broadly similar during the first four hours, despite differences in amplitude and peak timing. After 30 days, appreciable phase separation is visible in both backgrounds. In GR, the two trajectories have different finite-time FLI classifications, whereas at $\zeta=0.1$ both are classified as regular. Thus, the initial similarity and subsequent phase separation do not by themselves distinguish chaotic from regular motion.
 
    \begin{figure}[htbp]
    	\centering
    	\includegraphics[width=\textwidth]{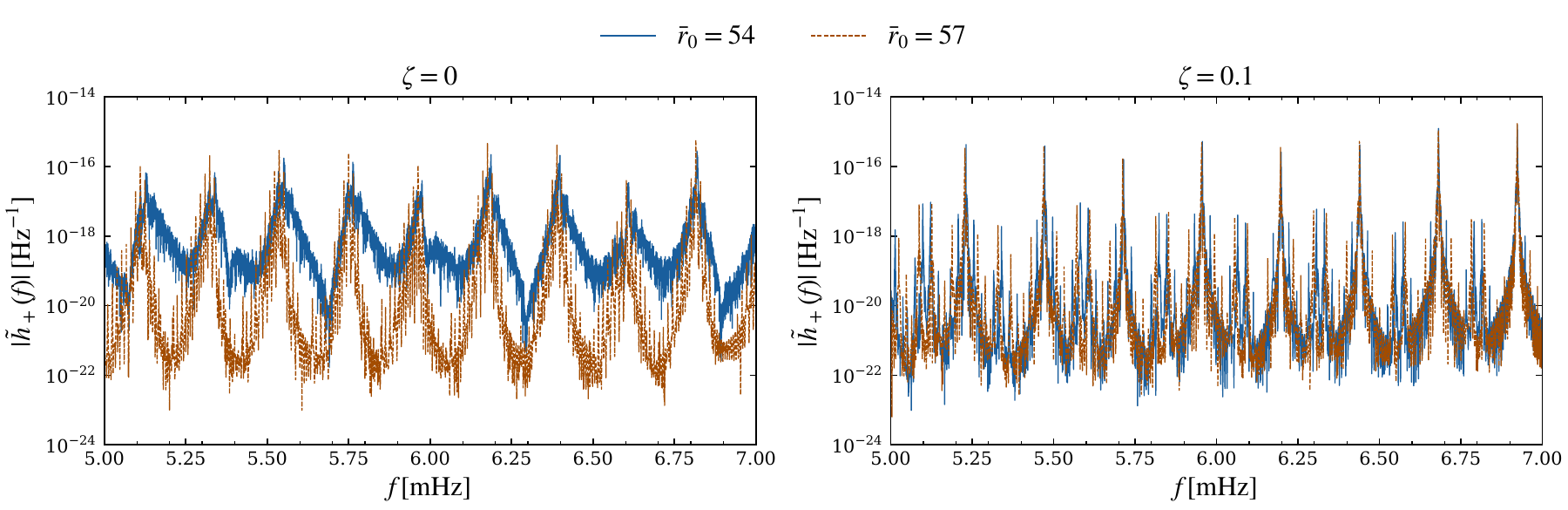}
    	\caption{Fourier amplitude spectra of $h_+$ for the trajectories shown in Figure~\ref{fig:waveforms-gr-zeta01}.}
    	\label{fig:spectra-gr-zeta01}
    \end{figure}
    
    To examine their spectral structure, we use the plus polarization $h_+(t)$ and define its Fourier transform as
    \begin{equation}
    	\tilde h_+(f)=\int_{-\infty}^{\infty}h_+(t)e^{-2\pi i f t}\,dt,
    	\label{eq:waveform-fourier}
    \end{equation}
    where $|\tilde h_+(f)|$ is the Fourier amplitude spectrum. In the numerical calculation, we use a common duration of $T=35$ days and a sampling rate of $0.075\,\mathrm{Hz}$. To reduce spectral leakage caused by the finite observation interval, we apply a Tukey window to each time series before computing its Fourier transform. The window parameter $\alpha=0.1$ specifies the fraction of the interval occupied by the cosine tapers, with $5\%$ at each end and unit weight over the central $90\%$. The spectra discussed below are those of these windowed signals. 
    
    Figure~\ref{fig:spectra-gr-zeta01} presents an enlarged view of the amplitude spectra in the $5$--$7\,\mathrm{mHz}$ band. In GR, the chaotic trajectory with $\bar r_0=54$ displays broader spectral features and higher amplitudes between the dominant peaks than the regular trajectory with $\bar r_0=57$. At $\zeta=0.1$, where both trajectories are classified as regular, their spectra show more similar structures. We therefore focus on the distribution of power around and between the spectral peaks, rather than on their frequency locations alone. Local spectral flatness provides a quantitative measure of this distribution and is unchanged by a uniform rescaling of the waveform amplitude.
    
    %%%%%%%%%%%%%%%%%%%%%%%%%%%%%%%%%%%%%%%%%%%%%%%%%%%%%%%%%%%%%%%%%%%%%%%%%%%%%%%%%%%%%%%%%%%%%%%%%%%%%%%%%%%%%%%%%%%%%%%
    \subsection{Local spectral flatness}
    \label{sec:spectral_flatness}
    %%%%%%%%%%%%%%%%%%%%%%%%%%%%%%%%%%%%%%%%%%%%%%%%%%%%%%%%%%%%%%%%%%%%%%%%%%%%%%%%%%%%%%%%%%%%%%%%%%%%%%%%%%%%%%%%%%%%%%%
    
    For this purpose, we define the spectral power at each frequency bin as $P_k=|\tilde h_+(f_k)|^2$. Within a sliding window $\mathcal W(f_c)$ centered at $f_c$, the local spectral flatness is \cite{YuanEtAl2026}
    \begin{equation}
    	\mathcal F(f_c)=\frac{\exp\left[\frac{1}{N_w}\sum_{k\in\mathcal W(f_c)}\ln P_k\right]}{\frac{1}{N_w}\sum_{k\in\mathcal W(f_c)}P_k},
    	\label{eq:local-spectral-flatness}
    \end{equation}
    where $N_w$ is the number of frequency bins in the window. This ratio compares the geometric and arithmetic means of the local spectral power. For a spectrum dominated by isolated sharp peaks, the low-power bins between the peaks suppress the geometric mean, yielding a small value of $\mathcal F$. When these intervals contain more spectral power, the distribution becomes more uniform and $\mathcal F$ generally increases. Equal power in all bins gives $\mathcal F=1$.
    
\begin{figure}[!htbp]
    	\centering
    	\includegraphics[width=\textwidth]{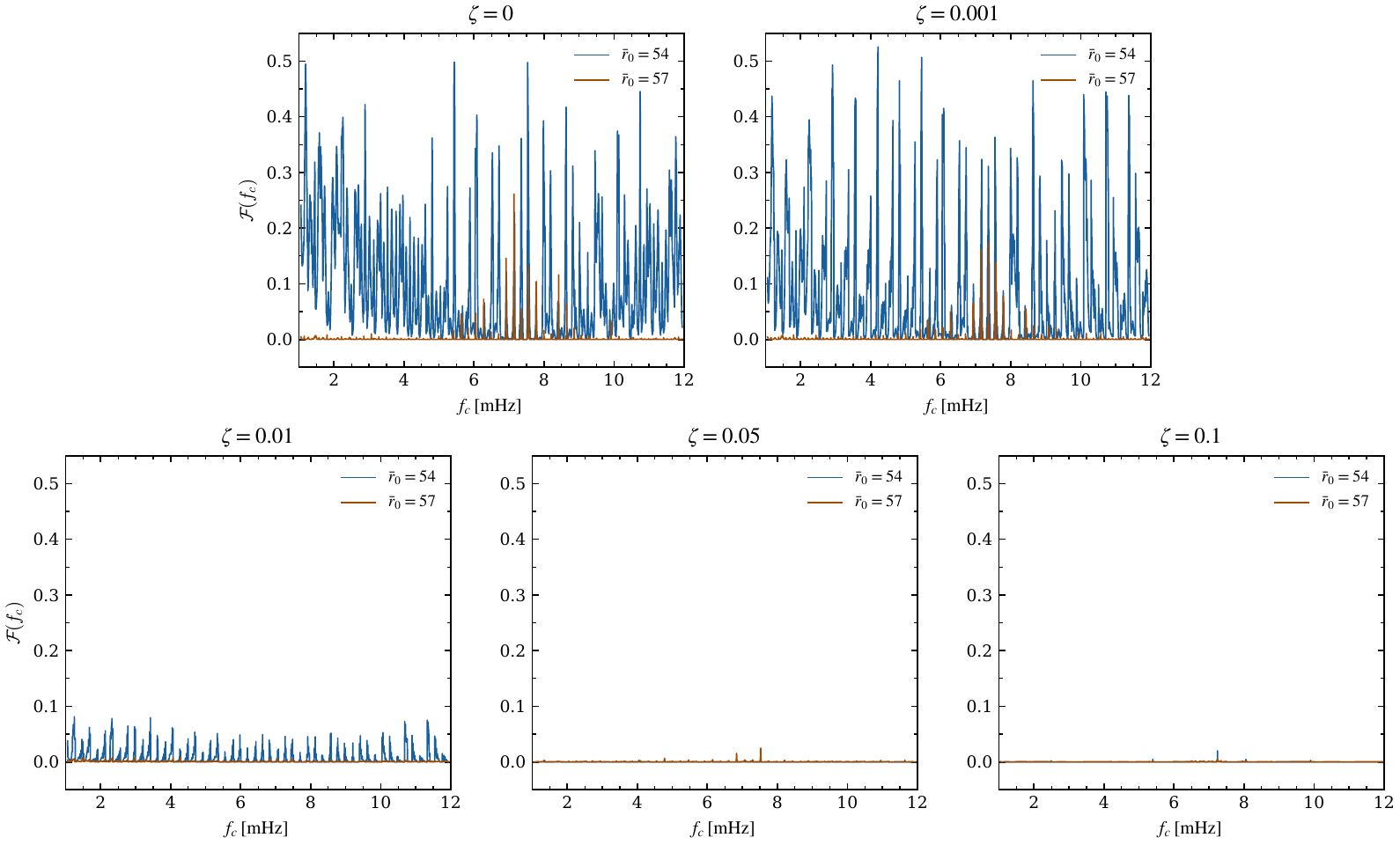}
    	\caption{Local spectral flatness of $h_+$ for the trajectories with $\bar r_0=54$ and $57$. The waveform duration, sampling rate, and Tukey window are the same as those used in Figure~\ref{fig:spectra-gr-zeta01}.}
    	\label{fig:local-spectral-flatness}
    	
    	\vspace{8pt}
    	
    	\captionof{table}{Mean local spectral flatness of the trajectories shown in Figure~\ref{fig:local-spectral-flatness}. The dynamical classifications follow the finite-time FLI criterion. The last column gives the ratio of the two mean values.}
    	\label{tab:mean-spectral-flatness}
    	\begingroup
    	\renewcommand{\arraystretch}{1.35}
    	\setlength{\tabcolsep}{10pt}
    	\begin{tabular}{cccc}
    		\hline\hline
    		$\zeta$ & $\overline{\mathcal F}\;(\bar r_0=54)$ & $\overline{\mathcal F}\;(\bar r_0=57)$ & $\overline{\mathcal F}_{54}/\overline{\mathcal F}_{57}$ \\
    		\hline
    		$0$       & $9.40\times10^{-2}$ (Chaotic) & $1.48\times10^{-3}$ (Regular) & $63.7$ \\
    		$10^{-3}$ & $7.68\times10^{-2}$ (Chaotic) & $1.37\times10^{-3}$ (Regular) & $56.1$ \\
    		$10^{-2}$ & $7.41\times10^{-3}$ (Chaotic) & $5.91\times10^{-4}$ (Regular) & $12.5$ \\
    		$0.05$    & $6.30\times10^{-5}$ (Regular) & $1.87\times10^{-4}$ (Regular) & $0.336$ \\
    		$0.1$     & $7.87\times10^{-5}$ (Regular) & $8.01\times10^{-5}$ (Regular) & $0.983$ \\
    		\hline\hline
    	\end{tabular}
    	\endgroup
    \end{figure}

    We use $N_w=321$, comprising the central bin and $160$ bins on either side, and move the window by one frequency bin at each step. Only windows fully contained within the $1$--$12\,\mathrm{mHz}$ band are included. To summarize the local results, we calculate the mean local spectral flatness,
    \begin{equation}
    	\overline{\mathcal F}=\frac{1}{N_c}\sum_{i=1}^{N_c}\mathcal F(f_{c,i}),
    	\label{eq:mean-local-spectral-flatness}
    \end{equation}
    where $i$ labels the successive sliding windows, $f_{c,i}$ is the center frequency of the $i$th window, and $N_c$ is the total number of retained windows.  Figure~\ref{fig:local-spectral-flatness} shows the frequency dependence of the local spectral flatness, and Table~\ref{tab:mean-spectral-flatness} lists the corresponding mean values.
    
    For $\zeta=0$, $10^{-3}$, and $10^{-2}$, the mean local spectral flatness of the chaotic trajectory with $\bar r_0=54$ is approximately one to two orders of magnitude larger than that of the regular trajectory with $\bar r_0=57$. Although their FLI classifications remain unchanged, the contrast decreases as $\zeta$ increases across these values. At $\zeta=0.05$ and $0.1$, both trajectories are classified as regular and have small mean flatness values. The value for $\bar r_0=54$ is about one-third of that for $\bar r_0=57$ at $\zeta=0.05$, while the two are nearly equal at $\zeta=0.1$. Thus, the pronounced flatness contrast between the chaotic and regular waveforms at small $\zeta$ becomes much weaker in the backgrounds where both trajectories are regular. The higher flatness of the selected chaotic waveform is consistent with its stronger inter-peak spectral content. Conversely, the low values at $\zeta=0.05$ and $0.1$ indicate more concentrated spectral power distributions under the adopted analysis settings.
    
    Within a given background at fixed $\zeta$, spectral flatness quantifies the spectral differences associated with the dynamical classifications of the selected trajectories. Comparisons across backgrounds also involve changes in orbital frequencies and harmonic amplitudes, which alter the power distribution within the fixed frequency band and sliding windows. Changes in flatness with $\zeta$ therefore cannot alone be interpreted as changes in the degree of chaos and should be considered alongside independent dynamical diagnostics such as the FLI.
    
    %%%%%%%%%%%%%%%%%%%%%%%%%%%%%%%%%%%%%%%%%%%%%%%%%%%%%%%%%%%%%%%%%%%%%%%%%%%%%%%%%%%%%%%%%%%%%%%%%%%%%%%%%%%%%%%%%%%%%%%
    \subsection{Waveform mismatch and degeneracy with initial radius}
    \label{sec:waveform_mismatch}
    %%%%%%%%%%%%%%%%%%%%%%%%%%%%%%%%%%%%%%%%%%%%%%%%%%%%%%%%%%%%%%%%%%%%%%%%%%%%%%%%%%%%%%%%%%%%%%%%%%%%%%%%%%%%%%%%%%%%%%%
    
    The background deformation can change the dynamical classification of a selected orbit family, raising the question of whether such changes provide useful information for distinguishing NLG from GR. We examine this possibility through waveform comparisons, with particular attention to degeneracy with orbital initial conditions. We first compare corresponding initial-radius labels across the two backgrounds, then allow the GR initial radius to vary to determine whether the resulting waveform differences can be reproduced within GR. 
    
    We choose $\zeta=10^{-6}$ to study waveform differences close to the Schwarzschild limit and examine whether a small background deformation produces changes that cannot be readily reproduced by varying the GR initial radius. The source parameters are fixed at $(M=10^6M_\odot,\ \mu/M=10^{-4},\ s=10^{-4}\mu M,\ J_z=3.8\mu M)$. We select seven NLG trajectories based on their finite-time FLI classifications. The trajectories with $\bar r_0=46$, $47$, and $54$ are chaotic in both backgrounds, while those with $\bar r_0=44$, $45$, and $49$ are regular in both. The trajectory with $\bar r_0=43$ is chaotic in GR but regular in NLG. The sample thus includes a case with a change in dynamical classification and controls whose classifications remain unchanged. For each selected NLG trajectory, we compare its waveform with GR waveforms generated at different initial radii. All GR waveforms use the same energy $E_{\mathrm h}(0)$ and the source parameters given above. Each NLG target is compared with the same set of GR waveforms, and the one with the smallest mismatch is selected as its best match.
    
    We quantify waveform agreement using the noise-weighted inner product
    \begin{equation}
    	(h_1|h_2)=4\,\mathrm{Re}\int_{0}^{f_{\max}}
    	\frac{\tilde h_1(f)\tilde h_2^*(f)}{S_n(f)}\,df,
    	\label{eq:mismatch-inner-product}
    \end{equation}
    where $S_n(f)$ is the one-sided noise power spectral density and the asterisk denotes complex conjugation. We adopt the LISA sensitivity model of Ref.~\cite{RobsonCornishLiu2019}, including instrumental noise and the Galactic confusion foreground. The normalized match is maximized over relative time and phase offsets,
    \begin{equation}
    	\mathcal{FF}
    	=
    	\max_{t_c,\phi_c}
    	\frac{
    		\left(h_1\middle|h_2e^{i(2\pi f t_c+\phi_c)}\right)
    	}{
    		\sqrt{(h_1|h_1)(h_2|h_2)}
    	},
    	\label{eq:maximized-match}
    \end{equation}
    where the phase factor acts on $\tilde h_2(f)$ in the frequency-domain inner product. Here $t_c$ is the relative time shift and $\phi_c$ is a constant waveform phase offset. The phase maximization is performed analytically by taking the absolute value of the complex overlap, followed by a search over time shifts. The mismatch is
    \begin{equation}
    	\mathcal M=1-\mathcal{FF}.
    	\label{eq:waveform-mismatch}
    \end{equation}
    A smaller mismatch indicates closer agreement after the time and phase adjustments.
    
    We use the plus polarization $h_+$ over a common duration of $35$ days, sampled at $0.125\,\mathrm{Hz}$. The inner product includes positive-frequency bins up to $f_{\max}=0.05\,\mathrm{Hz}$, with the zero-frequency bin omitted. A Tukey window with $\alpha=0.1$ is applied before the Fourier transform. The observer configuration is the same as in the preceding waveform comparisons, and all NLG targets and GR reference waveforms are processed with these settings.
    
    We construct the GR reference set by scanning and progressively refining the initial radius, with $\mathcal T$ denoting the set of initial-radius labels included. The initial set contains the 32 trajectories that complete the original scan, with $\bar r_0^{\mathrm{GR}}=34,35,36$ and $41,\ldots,69$. According to Eq.~\eqref{bar_r}, one label unit corresponds to a radial separation of $10^{-6}M$. For each NLG target, we first calculate the mismatch with the GR waveform carrying the same initial-radius label, denoted by $\mathcal M_{\mathrm{same}}$, and then search the reference set for a closer match. At the start of each refinement round, every NLG target is compared with all available GR reference waveforms. For each target, we first select the GR initial radius giving the smallest mismatch. We then select a second radius giving the smallest mismatch among those separated from the first by at least $10^{-6}M$. Additional points are sampled within $\pm10^{-6}M$ of both candidate radii. The points proposed by all seven targets are combined and duplicates are removed. Only previously uncalculated GR trajectories are integrated, and their waveforms are added to the common reference set. At the end of each round, the two candidate radii are updated using the matching results from all old and new reference waveforms. The sampling intervals in the three rounds are $5\times10^{-7}M$, $2.5\times10^{-7}M$, and $10^{-7}M$, corresponding to spacings of $0.5$, $0.25$, and $0.1$ in $\bar r_0^{\mathrm{GR}}$. Refinement is restricted to the candidate regions, while all previously calculated reference waveforms are retained and remain available for matching. The minimum mismatch for each NLG target over the reference set is defined as
    \begin{equation}
    	\mathcal M_{\mathcal T}(\bar r_0^{\mathrm{NLG}})
    	=
    	\min_{\bar r_0^{\mathrm{GR}}\in\mathcal T}
    	\mathcal M\!\left[
    	h_+^{\mathrm{NLG}}(\bar r_0^{\mathrm{NLG}}),
    	h_+^{\mathrm{GR}}(\bar r_0^{\mathrm{GR}})
    	\right].
    	\label{eq:radius-bank-mismatch}
    \end{equation}
    At each stage, the minimum is taken over the entire enlarged reference set. Comparing $\mathcal M_{\mathrm{same}}$ with $\mathcal M_{\mathcal T}$ shows how much the waveform agreement improves when the GR initial radius is adjusted.
      
    \begin{table}[htbp]
    	\centering
    	\caption{Mismatch between the selected NLG waveforms at $\zeta=10^{-6}$ and GR reference waveforms. The two classification columns refer to the NLG target and the GR trajectory with the same initial-radius label, respectively, using the finite-time FLI criterion. The remaining columns give the same-label mismatch, the best matching GR label in the final reference set, and the minimum mismatch.}
    	\label{tab:gr-nlg-mismatch}
    	\begingroup
    	\renewcommand{\arraystretch}{1.25}
    	\setlength{\tabcolsep}{8pt}
    	\begin{tabular}{cccccc}
    		\hline\hline
    		$\bar r_0^{\mathrm{NLG}}$ & NLG class & Same-label GR class & $\mathcal M_{\mathrm{same}}$ & Best-fit $\bar r_0^{\mathrm{GR}}$ & $\mathcal M_{\mathcal T}$ \\
    		\hline
    		43 & Regular & Chaotic & 0.34198 & 43.1 & 0.00242 \\
    		44 & Regular & Regular & 0.00249 & 44.1 & 0.00241 \\
    		45 & Regular & Regular & 0.00412 & 44.3 & 0.00356 \\
    		49 & Regular & Regular & 0.00249 & 49.1 & 0.00242 \\
    		\hline
    		46 & Chaotic & Chaotic & 0.38823 & 46.2 & 0.22738 \\
    		47 & Chaotic & Chaotic & 0.31453 & 46.1 & 0.12774 \\
    		54 & Chaotic & Chaotic & 0.66922 & 52.9 & 0.14322 \\
    		\hline\hline
    	\end{tabular}
    	\endgroup
    \end{table}
    
    Table~\ref{tab:gr-nlg-mismatch} shows that adjusting the GR initial radius can substantially change the comparison. The clearest example is the regular NLG target with $\bar r_0=43$. Its mismatch with the same-label GR waveform is approximately $0.342$, but the search reduces this to $0.00242$. Thus, the large same-label mismatch in this case mainly reflects the restriction to a particular GR counterpart. A nearby GR initial condition produces a much closer waveform. All four regular NLG targets have minimum mismatches between approximately $0.0024$ and $0.0036$. The three chaotic targets retain larger values, between $0.128$ and $0.227$. Within the searched GR family, the selected chaotic NLG waveforms are therefore harder to reproduce by adjusting the initial radius than the regular controls. This difference remains after every target has been compared with the same template bank.   
    
    The case with $\bar r_0=43$ shows that a change in dynamical classification does not necessarily produce a waveform difference that remains distinct after adjusting the GR initial radius. The selected chaotic NLG waveforms are harder to match within the searched GR family, but their larger residual mismatches do not by themselves establish an advantage for background identification. Sensitivity to initial conditions may also make chaotic signals harder to recover within the correct background.

    %%%%%%%%%%%%%%%%%%%%%%%%%%%%%%%%%%%%%%%%%%%%%%%%%%%%%%%%%%%%%%%%%%%%%%%%%%%%%%%%%%%%%%%%%%%%%%%%%%%%%%%%%%%%%%%%%%%%%%%
    \section{Conclusions}
    \label{sec:conclusions}
    %%%%%%%%%%%%%%%%%%%%%%%%%%%%%%%%%%%%%%%%%%%%%%%%%%%%%%%%%%%%%%%%%%%%%%%%%%%%%%%%%%%%%%%%%%%%%%%%%%%%%%%%%%%%%%%%%%%%%%%v
    
    We have studied how a nonlocal deformation of Schwarzschild spacetime affects spin-induced chaos and the associated approximate gravitational waveforms. The Poincar\'e sections and rotation curves remain broadly similar to those in GR for small deformations, while clearer changes appear as the deformation increases. These include changes in the resonant structures, smaller scattered regions, and smoother rotation curves. The FLI scans support a decreasing occurrence of detectable finite-time chaos within the chosen initial-condition family: among non-plunging trajectories, the fraction classified as chaotic falls from $37.5\%$ in GR to zero at the largest deformation considered. 
    
    The waveform analysis shows how these orbital differences appear in the radiation spectra. Similar early time-domain signals can correspond to different dynamical classifications, and later phase separation occurs for both regular and chaotic motion. For the selected neighboring trajectories, the chaotic waveform has a mean local spectral flatness approximately one to two orders of magnitude larger than the regular waveform. This contrast becomes much smaller in backgrounds where both trajectories are classified as regular. Spectral flatness thus provides a quantitative counterpart to the independently determined orbital classifications. Its variation across backgrounds, however, also depends on changes in orbital frequencies and harmonic amplitudes, so it cannot be used alone to measure changes in the degree of chaos.
    
    We also examined whether these dynamical changes help distinguish NLG from GR. In the selected case that changes from chaotic motion in GR to regular motion in NLG, adjusting the GR initial radius reduces the waveform mismatch by more than two orders of magnitude. The classification change therefore does not, by itself, leave a waveform difference that is robust against this initial-condition adjustment. The selected chaotic NLG waveforms retain minimum mismatches roughly one to two orders of magnitude larger than those from regular orbits, making them harder to match within the searched GR family. This suggests a possible sensitivity to the background geometry, but does not yet demonstrate an advantage for identifying it. Sensitivity to initial conditions may also make chaotic signals harder to recover within the correct background.
    
    From a physical perspective, both background deformations and changes in orbital initial conditions can alter orbital frequencies, long-term phase accumulation, and radiation spectra. A strong waveform response to a background deformation therefore does not necessarily provide a unique signature of the background geometry. Its usefulness for distinguishing gravitational backgrounds depends on whether the resulting waveform features can be separated from changes caused by orbital initial conditions and other source parameters.
    
    The present study uses conservative pole--dipole dynamics and approximate mass-quadrupole numerical kludge waveforms. An important extension is to include the secondary's spin-induced quadrupole in the equations of motion and examine whether it changes the resonances and chaos found here. Studies of quadrupolar test bodies in Kerr spacetime show that these additional couplings can affect the integrability properties of the dynamics \cite{CompereDruartVines2023,Ramond2025}. Another direction is to include radiation reaction and study resonance crossings, the time spent within chaotic layers, and the resulting frequency evolution and phase accumulation \cite{LukesGerakopoulosWitzany2022}. Related studies of inspirals into rotating boson stars have examined resonance crossings and their waveform signatures using approximate radiation fluxes \cite{DestounisEtAl2023}. Such an extension would test whether the structures found here in the conservative approximation leave identifiable waveform features during dissipative EMRI evolution.

	\acknowledgments
	
	This work is supported by the National Natural Science Foundation of China (NSFC) with Grant No. 12275087.
	
	\bibliographystyle{jhep}
	\bibliography{ref}
	
\end{document}